\documentclass[referee, Numbered, sn-mathphys]{sn-jnl}% referee option is meant for double line spacing

\usepackage{graphicx}%
\usepackage{multirow}%
\usepackage{amsmath,amssymb,amsfonts}%
\usepackage{amsthm}%
\usepackage{mathrsfs}%
\usepackage[title]{appendix}%
\usepackage{xcolor}%
\usepackage{textcomp}%
\usepackage{manyfoot}%
\usepackage{booktabs}%
\usepackage{algorithm}%
\usepackage{algorithmicx}%
\usepackage{algpseudocode}%
\usepackage{listings}%
\usepackage[utf8]{inputenc} %unicode support
\usepackage{longtable}
\usepackage{pifont}
\usepackage{mathtools}
\usepackage{caption}
\usepackage{subcaption}
\usepackage{makecell} % force line break in the cell header
\usepackage{graphicx}
\usepackage{placeins} % figure appears in the section in which it's declared
\usepackage{afterpage} % used to place the second subfigure on the next page
\usepackage{verbatim}
\usepackage{xr}
\usepackage[switch, displaymath, mathlines]{lineno}
\usepackage{bbm}
\usepackage{bm}
\usepackage{siunitx}
\usepackage{array}
\usepackage{lipsum} % For generating dummy text
\usepackage{titlesec} % For advanced section formatting
\usepackage{float}
\usepackage{csquotes}

\newcolumntype{R}[1]{>{\raggedleft\arraybackslash}p{#1}}   % Rechts ausgerichtet

\begin{document}

\title[Individual Time Effects in Cohort Stepped Wedge Trials: A Simulation Study]{Guidance for Addressing Individual Time Effects in Cohort Stepped Wedge Cluster Randomized Trials: A Simulation Study}


\author*[1]{\fnm{Jale} \sur{Basten}}\email{basten@amib.ruhr-uni-bochum.de}

\author[2,3]{\fnm{Katja} \sur{Ickstadt}}\email{ickstadt@statistik.tu-dortmund.de}

\author[1]{\fnm{Nina} \sur{Timmesfeld}}\email{timmesfeld@amib.ruhr-uni-bochum.de}

\affil[1]{\orgdiv{Department of Medical Informatics, Biometry and Epidemiology}, \orgname{Ruhr-University of Bochum}, \orgaddress{\street{Universitätsstrasse 142}, \city{Bochum}, \postcode{44799}, \state{North-Rhine Westphalia}, \country{Germany}}}

\affil[2]{\orgdiv{Faculty of Statistics}, \orgname{TU Dortmund University}, \orgaddress{\street{Vogelpothsweg 78}, \city{Dortmund}, \postcode{44227}, \state{North-Rhine Westphalia}, \country{Germany}}}

\affil[3]{\orgdiv{Lamarr-Institute for Machine Learning and Artificial Intelligence}, \orgaddress{\street{Joseph-von-Fraunhofer-Strasse 25}, \city{Dortmund}, \postcode{44227}, \state{North-Rhine Westphalia}, \country{Germany}}}

\abstract{\textbf{Background:} Stepped wedge cluster randomized trials (SW-CRTs) involve sequential measurements within clusters over multiple time periods. Initially, all clusters begin in the control condition before crossing over to the intervention on a staggered schedule, resulting in an unbalanced allocation of observations to the intervention groups over time. In cohort designs, where the same participants are observed repeatedly, not only secular trends and cluster-level changes but also individual-level changes (e.g., aging) must be considered.

\textbf{Methods:} Inspired by an SW-CRT evaluating a geriatric care intervention, we performed a Monte Carlo simulation to analyze the influence of different time effects on the estimation of the intervention effect in cohort SW-CRTs. To evaluate how time-varying covariates - both linear and nonlinear - affect model performance, we compared four linear mixed models that use different adjustment strategies. All models included random intercepts for clustering and repeated measurements of individual-level outcomes. For each simulated dataset and analysis model, we recorded the estimated fixed intervention effects and their corresponding model-based standard errors, derived from models both without and with cluster-robust variance estimators (CRVEs), specifically CR0VE, CR2VE, and CR3VE.

\textbf{Results:} Models incorporating fixed categorical time effects, a fixed intervention effect, and two random intercepts for clustering and repeated measurements of individuals provided unbiased estimates of the intervention effect in both closed and open cohort SW-CRTs, even when time-varying covariates were unmeasured or their functional form was unknown. Fixed categorical time effects captured temporal cohort changes, while random individual intercepts accounted for differences in participants' baseline characteristics. However, these baseline differences can cause large, non-normally distributed random individual intercepts. Additionally, unknown time-varying covariates with nonlinear effects may induce residual heteroscedasticity. CRVEs provide reliable standard errors for the intervention effect, even when there are complex residual dependencies, thereby controlling the type I error rate.

\textbf{Conclusions:} Our simulation study is the first to assess individual-level changes over time in cohort SW-CRTs. Linear mixed models incorporating fixed categorical time effects and random cluster and individual intercepts yield unbiased intervention effect estimates across the scenarios considered. However, valid inference depends on appropriate variance estimation. In settings where time-varying covariate effects are nonlinear or unknown, the CR3VE (Mancl \& DeRouen) demonstrated more robust confidence interval coverage and type I error control, suggesting that it may represent a pragmatic default approach, albeit potentially at the expense of wider confidence intervals and some loss of statistical power, particularly in small-sample settings.
}




\maketitle

\section{Background}
\label{sec:background}
In cluster randomized trials (CRTs), the clusters - such as hospitals, medical practices or geographical units - are the units of randomization, rather than individuals~\cite{Murray2004}. This design is particularly advantageous when interventions are naturally applied at the cluster level or when it is necessary to prevent intervention contamination~\cite{Torgerson2001}. A key consequence of this approach is that individuals within the same cluster tend to be more similar to one another than individuals from different clusters. This similarity is quantified by the intracluster correlation coefficient (ICC)~\cite{Kerry1998}. Due to their more complex correlation structure, CRTs typically require larger sample sizes than individually randomized trials~\cite{Turner2017_PART2}.
\par
\begin{figure}[!htb]
 \centerline{\includegraphics[width=240pt,height=10pc]{"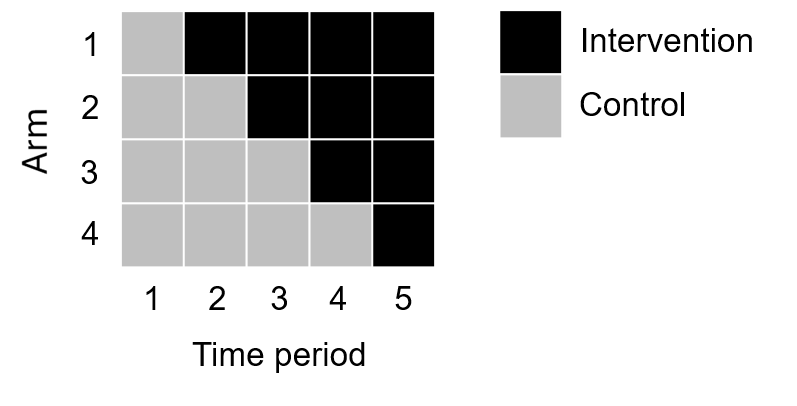"}}
 \caption{Schematic representation of a standard SW-CRT with four arms/steps.\label{fig:SWD}}
\end{figure}

However, longitudinal CRT designs, such as the stepped wedge cluster randomized trial (SW-CRT), can result in a reduced required sample size~\cite{Woertman2013, Hooper2016, Hemming2015_2}. The defining feature of SW-CRTs is the unidirectional crossover of clusters from the control to intervention conditions on a staggered schedule (Figure~\ref{fig:SWD}). The initial time point corresponds to a baseline measurement, during which none of the clusters receive the intervention of interest (pre-rollout period). This is followed by the rollout phase, during which several clusters cross over from the control condition to the intervention condition at distinct time points, referred to as steps. Clusters that switch simultaneously form a randomization group, or trial arm. In each step, one trial arm begins the intervention while the remaining clusters continue in the control or intervention condition. This process continues until all clusters have received the intervention, entering the post-rollout period~\cite{Hussey2007}. This results in an unbalanced allocation of observations to the intervention groups over time, with control observations collected before intervention observations.

\FloatBarrier

SW-CRTs can facilitate the gradual implementation of an intervention across multiple settings, making it logistically feasible for complex interventions, involving multiple interacting components and requiring behavioral changes from various stakeholders~\cite{Skivington2021}. Examples include the rollout of a new vaccination program across multiple communities~\cite{MoralesCampos2025}, the sequential training of healthcare workers in a new procedure across several hospitals~\cite{Vidal2025}, or the phased introduction of a new educational curriculum across schools to monitor its impact~\cite{Quach2024}.
\par
Unlike conventional CRTs, SW-CRTs have the unique advantage of exposing each cluster to both the control and the intervention conditions. This design enables both within-cluster and between-cluster comparisons, often referred to as horizontal and vertical comparisons, respectively. Vertical approaches compare the outcomes of clusters in the intervention condition with the outcomes of clusters in the control condition within the same time period, while horizontal approaches compare outcomes before and after crossover to the intervention condition. However, horizontal comparisons can be confounded by time-related effects~\cite{Hargreaves2015}. Consequently, time-related effects pose a well-documented challenge in SW-CRTs~\cite{Hemming2015_2}.
\par
Depending on the intervention and participant recruitment, different approaches can be chosen for the SW-CRT, drawing from the terminology of Feldman and McKinlay (1994)~\cite{Feldman1994}. In a cross-sectional design, participants contribute a single outcome during the study, meaning different participants are considered at each step. Conversely, in closed cohort designs, all participants are identified at the beginning of the trial and participate throughout its duration. Additionally, Copas et al.~\cite{Copas2015} introduced a third option: the open cohort design. In this approach, a substantial number of participants are included and participate from the beginning, but some may drop out during the trial, while others may become eligible and be exposed for a certain period. For instance, in a medical context, the cohort design typically includes chronically ill patients who are the target group of the intervention over several time periods. In contrast, a cross-sectional design observes newly ill patients who are no longer part of the target population soon after the intervention. While the methodological focus of SW-CRTs has been on cross-sectional designs, as originally proposed by Hussey and Hughes (2007)~\cite{Hussey2007}, the majority of SW-CRTs in practice are studies with closed or open cohort designs~\cite{Martin2016}.
\par
A commonly used analytical model for SW-CRTs is the linear mixed model (LMM)~\cite{Li2020_Overview}. For cross-sectional designs, Hussey and Hughes proposed a basic model in 2007~\cite{Hussey2007}, which was extended for cohort designs by Baio et al. in 2015~\cite{Baio2015}. This basic model is a random intercept model, accounting for within-cluster and within-individual correlations. In our simulation study, we focus on random intercept models for clustering and repeated measurements of individual-level outcomes. In the literature, more sophisticated structures have been proposed, including random cluster-by-period effects, discrete-time decay, and, more recently, the random intervention structure~\cite{Hooper2016, Girling2016, Kasza2019.2, Hughes2015, Hemming2017}. For binary or count outcomes, generalized linear mixed models (GLMMs) with appropriate link functions are typically used. Although rarely applied in this context, fully nonlinear mixed effects models offer further flexibility~\cite{Pinheiro2002}.
\par
Like most longitudinal trial designs, SW-CRTs can be influenced by external factors - such as new public health policies, seasonal fluctuations, or COVID-19-related lockdowns~\cite{BrünnBasten2024} - that may affect outcomes. Because the proportion of clusters receiving the intervention increases with time, these external trends align with intervention rollout and can confound the estimated intervention effect if not appropriately accounted for (i.e., confounding by calendar time)~\cite{Hemming2015_2}. Simulation studies have shown that such calendar time effects must be taken into account to avoid highly biased estimates of the intervention effect and type I and type II errors~\cite{Rennert2021, Nickless2018}.
\par
The calendar time effect usually affects all clusters in the same way. However, they may also influence outcomes differently across clusters at different times. For example, depending on its location, a primary care practice may be affected by a flu epidemic at different times and to varying degrees. Rennert et al. (2021)~\cite{Rennert2021} demonstrated that incorporating fixed intervention-by-time interactions with an unstructured covariance for intervention-by-cluster-by-time random effects can reduce bias in settings where external factors differentially impact intervention and control clusters.
\par
According to the Consolidated Standards of Reporting Trials (CONSORT) extension for SW-CRTs~\cite{Hemming2019}, time effects are one of the most important potential confounders of SW-CRTs, leading to bias~\cite{Barker2016}, and therefore require special attention at both the design and analysis stages. The guideline indicates that calendar time effects should be considered in the analysis of SW-CRTs, regardless of their statistical significance. Otherwise, there is a risk that the estimate of the intervention effect will be biased, and an intervention may be falsely declared ineffective when it is effective, or effective when it is ineffective. It is therefore essential to report whether and how time effects were taken into account~\cite{Hemming2019}.
\par
In addition to trends in calendar time, structural changes at the cluster level can also occur. For example, staffing situations may fluctuate over time. Furthermore, cohort designs face specific challenges arising from individual-level changes. For example, aging can lead to natural decline, and life events during the study period can affect outcomes. We refer to these fluctuations as individual time effects.
\par
To the best of our knowledge, no studies have yet examined how to address individual time effects, a distinctive characteristic of cohort SW-CRTs. The analysis of cohort SW-CRTs is more complex than that of cross-sectional SW-CRTs, primarily due to their more complex correlation structures and the need for careful consideration of individual time effects. Therefore, it is essential to clearly distinguish between cohort and cross-sectional designs in trial reporting~\cite{Barker2016, Martin2016}.
\par
Certain forms of individual time effects, such as nonlinear time-varying covariates, may induce heteroscedasticity of residuals in LMMs, which may require cluster-robust variance to control type I error rates appropriately. Schielzeth et al. (2020)~\cite{Schielzeth2020} and Wang et al. (2024)~\cite{Wang2024b} demonstrated that the LMM is robust against violations in distributional assumptions of residual and random effects. However, recent studies by Hui et al. (2021)~\cite{Hui2021}, Voldal et al. (2022)~\cite{Voldal2022}, and Ouyang et al. (2024)~\cite{Ouyang2024} have shown that misspecification can substantially affect statistical inference.
\par
Given the limited guidance available to trialists, we conduct a Monte Carlo simulation study focusing on time-varying covariates and informative attrition as key sources of individual time effects in cohort SW-CRTs. We evaluate mixed models with different covariate adjustment strategies and assess their impact on bias and statistical inference. In addition, we investigate the performance of cluster-robust variance estimators (CRVEs) in maintaining valid statistical inferences across a range of realistic scenarios.
\par
This article is structured as follows: Section~\ref{sec:motivation} presents a motivating example to contextualize the study. In Section~\ref{sec:methods}, we detail the methods and notation used in the simulation study, including the data-generating process, simulation parameters, analysis models, and performance measures. Section~\ref{sec:results} shows the results of the simulation study, which examines the performance of various covariate adjustment strategies and cluster-robust variance estimation techniques for cohort SW-CRTs across different scenarios. Section~\ref{sec:discussion} provides an in-depth discussion of the results, and Section~\ref{sec:conclusion} offers practical guidance on addressing individual time effects.

\section{Motivating example: \enquote{frail older Adults: Care in Transition} (ACT) trial}
\label{sec:motivation}
In cohort designs, study participants are typically chronically ill and do not fully recover, thus remain part of the target population throughout the study period. Consequently, frail older adults are often included in cohort SW-CRTs. An example of such a trial is the \enquote{frail older Adults: Care in Transition} (ACT) trial, registered in the Netherlands National Trial Register (NTR2160)~\cite{Muntinga2012, Hoogendijk2015}.
\par
The ACT trial evaluated the effectiveness of a geriatric care model on the quality of life of community-dwelling frail older adults and ran over a 24-month period. Quality of life was assessed using the 12-item Short Form questionnaire (SF-12), which comprises two domains: a mental health component score (MCS) and a physical health component score (PCS). In the ACT trial, only the MCS was utilized, and an increase over time was observed~\cite{Twisk2016}. The increase in the MCS may be attributable to several causes or combinations thereof:
\begin{enumerate}
  \item The intervention may have influenced the MCS, and due to the study design the proportion of observations in the intervention condition have increased (intervention effect).
  \item Secular trends may have led to a change in mental health independent of the intervention (calendar time effect).
  \item The study cohort may have changed over time due to aging or participant attrition, which could affect the MCS independently of the intervention (individual time effect).
\end{enumerate}
All three causes may have influenced the MCS in the same or different directions. For instance, the intervention may have been effective, leading to improved mental health. Alternatively, the observed increase in the MCS could also be attributed to secular trends that are independent of the intervention. These trends could include increased societal affluence, enhanced leisure activities for the elderly, and other factors. Improvements in outcomes due to external factors have been previously described using the term \enquote{rising tide}~\cite{Bion2013, Rennert2021}. Building on this metaphor, individual aging can be viewed as a \enquote{separate current}, representing changes in participants' health trajectories that unfold independently of these population-level developments and may even counteract them. Accordingly, as participants age during the study period, they may experience increasing frailty or significant life changes that adversely affect mental health, such as serious illness or bereavement. In this way, the \enquote{separate current} of aging may exert a negative influence on mental health, even as the intervention and secular trends exert positive effects. However, if we consider an open cohort, it is possible that the study cohort will become younger and healthier. Alternatively, the cohort's age and health status may remain constant over the study period as older, frailer participants die and younger, healthier participants join the study. To estimate the intervention effect without bias, adjustments for calendar time effects and individual time effects are necessary.
\par
In the ACT trial, an increase in the outcome variable over time was observed across the entire population, independent of the intervention. Consequently, the unadjusted results demonstrated a strong positive intervention effect. This observed increase over time was incorrectly attributed to the intervention due to the greater proportion of participants in the intervention condition towards the end of the trial. However, the significant intervention effect observed in the unadjusted analyses disappeared upon adjustment for time~\cite{Twisk2016}.
\par
In addition to calendar time effects, individual time effects may arise from both within-individual changes (e.g., aging or disease progression) and changes in cohort composition over time due to attrition or recruitment. The authors acknowledged that frailty was only measured at baseline and not during follow-up. Additionally, up to 18\% of participants were lost to follow-up between consecutive measurements. This attrition is unlikely to be missing at random, as deceased participants, for example, could not provide further follow-up data. Therefore, it is reasonable to assume that the baseline cohort, in which all participants were in the control condition, cannot be directly compared with the cohort at the end of the study, in which all participants were in the intervention condition.
\par
Having identified the various factors that could influence the MCS within the ACT trial, we now address the key question: How should calendar and individual time effects be modeled in the analysis of cohort SW-CRTs to obtain unbiased intervention effect estimates and valid statistical inference?

\section{Simulation study methods}
\label{sec:methods}
This chapter outlines the methods employed in our Monte Carlo simulation study, wich was conducted using R version 4.5.3~\cite{RCoreTeam2021}.

\subsection{Data-generating process}
We implemented a standard SW-CRT design, incorporating both closed and open cohort measurements. This design assumes $I$ clusters ($i =1,\ldots,I$) and $J+1$ time periods ($j=1,\ldots,J+1$; these are $J$ steps), each of equal duration of one year (Figure~\ref{fig:SWD}). At each step, a proportion of $I/J$ clusters switch from the control period to the intervention period. Participants within each cluster-period cell are indexed by $k$, where $k = 1,..., K_{ij}$. In the cohort design, the same individuals are followed over time within each cluster, resulting in a constant cluster size, i.e., $K_{i1}=K_{i2}=...=K_{i(J+1)} \forall$ $i$ $\in$ $\{1,...,I\}$. In the open cohort design, we assume consistent attrition rates across all clusters and time periods.
\par
For data generation, we employ a basic model with only main fixed effects and random intercepts. Let $y_{ijk}$ denote the continuous outcome for participant $k$ at time period $j$ in cluster $i$. We assume that individual variability in the observed outcome data $y_{ijk}$ is described by a normal distribution with a cluster-specific, individual-specific, and time-specific mean $\mu_{ijk}$ and a variance $\sigma^2_e$, conditional on both the random cluster and the random individual intercept.
\par
Further, $\mu_{ijk}$ depends on a continuous time-varying covariate, hence the linear predictor was calculated as follows:
\begin{align*}
  \mu_{ijk} &= \mu + c_i + d_{ik} + \theta \times x_{ij} + \beta_{con} \times \left(a_{ijk}\right)^p,
\end{align*}
where notation is defined in Table~\ref{tab:notation_core} and parameter values are given in Table~\ref{tab:parameter}.

\begin{table}[!htb]
\caption{Core notation used throughout the simulation study.}\label{tab:notation_core}
\centering
\begin{tabular}{p{2.2cm}p{11.5cm}}
\toprule
\textbf{Notation} & \textbf{Definition} \\
\midrule
$i = 1,\ldots,I$ & Cluster index \\
$j = 1,\ldots,J+1$ & Time period index \\
$k = 1,\ldots,K_{ij}$ & Individual index within cluster $i$ at time $j$ \\
$y_{ijk}$ & Outcome for individual $k$ at time period $j$ in cluster $i$ \\
$\mu$ & Intercept of the LMM \\
$c_i$ & Random cluster intercept (between-cluster heterogeneity) \\
$d_{ik}$ & Random individual intercept (within-individual correlation) \\
$\theta$ & Intervention effect \\
$x_{ij}$ & Binary indicator of intervention \\
$\beta_{con}$ &  Effect of the time-varying covariate \\
$a_{ijk}$ & Value for the continuous time-varying covariate of joining participant $k$ at time period $j$ in cluster $i$ \\
\bottomrule
\end{tabular}
\end{table}
\FloatBarrier

\subsection{Simulation parameters}
We generated datasets for four scenarios: a closed cohort with a linear covariate effect, a closed cohort with a nonlinear covariate effect, an open cohort with a linear covariate effect, and an open cohort with a nonlinear covariate effect. Additionally, we used 12 parameter settings that varied by three cluster counts ($I = 8, 16, 32$), two step counts ($J = 4, 8$), and two cluster-period sizes ($K_{ij}= 10, 100$). Our choice was guided by empirical evidence from Nevins et al. (2024)~\cite{Nevins2024}, who reported substantial variation in SW-CRT designs (median 11 clusters, IQR 8-18; median 5 periods, IQR 4-7), as well as by the ACT trial~\cite{Hoogendijk2015}, introduced in Section~\ref{sec:motivation}, which included 35 clusters over four steps. The intervention effect ($\theta$) was varied across simulation runs (0, 0.25, 0.5, 0.75, 1) to assess both type I error (under $\theta = 0$) and statistical power. The time-varying covariate (e.g., age) was specified with either a linear ($p=1$; $\beta_{con} = -2$) or a nonlinear ($p=4$; $\beta_{con}=-2/\left(\frac{1}{32}\right)^4$) functional form. The nonlinear effect was scaled to preserve a comparable outcome range while clearly deviating from linearity. To increase sensitivity to individual time effects, the covariate was assumed to change by one unit per time period, corresponding to a faster progression than in the ACT trial (which used six-month intervals). The covariate effect size was chosen to be sufficiently strong to allow clear assessment of its impact on estimation and inference. Random effects were specified as cluster and individual intercepts with variances $\sigma_c^2 = 10$ and $\sigma_d^2 = 10$. The residual variance was $\sigma_e^2 = 20$. This yields a moderate within-individual ICC of 0.5, consistent with other SW-CRT settings~\cite{Li2020_Overview}. The open cohort SW-CRT was simulated with a 15\% per-period attrition rate with higher dropout among older individuals, reflecting non-random loss to follow-up in frail populations such as the ACT trial. New, on average younger, participants are added at each time point. A summary of the notation and simulated data scenarios is provided in Table~\ref{tab:parameter}.

\begin{sidewaystable}
\caption{Summary of notation and simulation study data scenarios.}\label{tab:parameter}
\begin{tabular*}{\textheight}{@{\extracolsep\fill}p{1.2cm}p{6cm}p{10.8cm}}
\toprule%
\textbf{Notation} &\textbf{Parameter description} &\textbf{Value and further description} \\
\midrule
$I$ & Number of clusters & 8, 16, 32\\
$J$ & Number of steps & 4 or 8. At each step $I/J$ clusters switch from the control to the intervention.\\
$J+1$ & Number of time periods & 5 or 9. The study duration is always one period longer than the number of steps because all clusters are observed under control conditions during the pre-rollout period.\\
$K_{ij}$ & Cluster size & 10 or 100 individuals in each cluster and time period in closed and open cohort design.
\\
$y_{ijk}$ & See Table~\ref{tab:notation_core} & $y_{ijk} \sim \mathcal{N}\left(\mu_{ijk}, \sigma^2_e\right)$ \\
$x_{ij}$ &  See Table~\ref{tab:notation_core} & $x_{ij} = \begin{cases} 0, \text{ if cluster $i$ is under control condition at time period $j$} & \\ 1, \text{ if cluster $i$ is under intervention condition at time period $j$} \end{cases}$\\
$l$ & Length of a step & one time unit, i.e., one time unit elapses from one crossover to intervention to the next.\\
$\mu$ & See Table~\ref{tab:notation_core} & 100\\
 & \\
\multicolumn{3}{p{18.2cm}}{\textbf{Fixed effects parameterisations}}\\
$\theta$ & Intervention effect & 0, 0.25, 0.5, 0.75, 1 common to all clusters. Furthermore, we assume that the intervention effect remains consistent over time and that there are no learning or weakening effects. \\
$(a_{ijk})^p$ & See Table~\ref{tab:notation_core} & For the pre-rollout period ($j=1$), the covariate $a_{i1k}$ follows a uniform distribution, $a_{i1k} \sim \text{Uniform}(18, 102)$ (e.g., participant’s age). We assume that this covariate increases by one unit (e.g., one year of age) at each subsequent step. The effect of this covariate is modeled as either linear, with an exponent $p=1$ in the linear predictor, or nonlinear, with an exponent $p=4$.\\
$\beta_{con}$ &  See Table~\ref{tab:notation_core} & -2 (linear), $-2/\left(\frac{1}{32}\right)^4$ (nonlinear). This parameter value for the nonlinear effect was chosen to maintain the same outcome range as in the linear case, while producing a clear nonlinear effect of the covariate.\\
\multicolumn{3}{p{18.2cm}}{\textbf{Random
 effects parameterisations}}\\
$c_i$ & See Table~\ref{tab:notation_core} & $c_i \sim \mathcal{N}(0, \sigma^2_c)$\\
$\sigma_c^2$ & Between-cluster variance & 10\\
$d_{ik}$ & See Table~\ref{tab:notation_core} & $d_{ik} \sim \mathcal{N}(0, \sigma^2_d)$\\
$\sigma^2_d$ & Between-individual variance & 10 \\
$\sigma^2_e$ & Random error variance & 20\\
$\rho_a$ & Within-individual ICC following Li et al. (2018)~\cite{Li2020_Overview} & $\rho_a = \frac{\sigma_c^2+\sigma_d^2}{\sigma_c^2+\sigma_d^2+\sigma^2_e} = 0.5$. Correlation between two repeated measurements from the same individual.\\
 & \\
\multicolumn{3}{p{18.2cm}}{\textbf{Open cohort parameterisations}}\\
$\chi(j,j+1)$ & Attrition rate between period $j$ and $j+1$ & 15\%\\
$(a_{ijk})^p$ & Value for the continuous time-varying covariate of joining participant $k$ at time period $j$ in cluster $i$ & For the pre-rollout period ($j=1$), the covariate $a_{i1k}$ follows a uniform distribution, $a_{i1k} \sim \text{Uniform}(18, 96)$ (e.g., participant’s age). We assume that this covariate increases by one unit (e.g., one year of age) at each subsequent step. The effect of this covariate is modeled as either linear, with an exponent $p=1$ in the linear predictor, or nonlinear, with an exponent $p=4$.\\
\botrule
\end{tabular*}
\end{sidewaystable}
\FloatBarrier

\subsection{Data analysis}
\label{subsec:AnalysisModels}
Each simulated dataset was analyzed using LMMs with restricted maximum likelihood (REML) estimation via the R package \texttt{lmerTest}~\cite{Kuznetsova2017}.
\par
We employed four different LMMs to account for a time-varying covariate, all including two random intercepts to account for clustering and within-individual correlation, as well as an intervention effect. In all models, the outcome $y_{ijk}$ has a mean $\mu_{ijk}$ and variance $\sigma^2_e$, conditional on the random cluster and random individual intercepts.
\par
Our basic model with fixed categorical time effects describes the mean of the outcome as follows:
\begin{align*}
  \mu_{ijk} = \mu + c_i + d_{ik} + \theta \times x_{ij} + \beta_{j},
\end{align*}
where $\beta_j$ corresponds to time period $j$ ($j \in 1,..., J+1, \beta_1 = 0$ for identifiability) and represents the estimated fixed categorical time effects. Our target estimand is the coefficient of the binary intervention indicator from the true LMM, denoted by $\theta$. In this framework, $\theta$ represents a conditional (cluster-specific) intervention effect. Under a linear model with an identity link, this coincides with the marginal (population-averaged) effect~\cite{Fitzmaurice2011, Kahan2022}. In the following, estimates are denoted with a hat symbol, as in $\hat{\theta}$. Table~\ref{tab:mixed_models} provides an overview of the LMM equations.

\begin{table}[!htb]
\caption{Summary of analysis models.}
\label{tab:mixed_models}
\centering
\begin{tabular}{p{0.15cm}p{5.5cm}p{6cm}}
    \toprule
    \textbf{Eq.} & \textbf{Description of adjustment} &\textbf{Parameter} \\
    \midrule
    1 & Time-updated linear covariate adjustment model & $\mu_{ijk} = \mu + c_i + d_{ik} + \theta \times x_{ij} + \beta_{con} \times a_{ijk}$\\
    2 & Fixed time effects model & $\mu_{ijk} = \mu + c_i + d_{ik} + \theta \times x_{ij} + \beta_{j}$\\
    3 & Fixed time effects and baseline covariate adjustment model & $\mu_{ijk} = \mu + c_i + d_{ik} + \theta \times x_{ij} + \beta_{j} +  \beta_{con} \times a_{i0k}$\\
    4 & Fixed time effects and time-updated covariate adjustment model & $\mu_{ijk} = \mu + c_i + d_{ik} + \theta \times x_{ij} + \beta_{j} +  \beta_{con} \times a_{ijk}$\\
    \botrule
\end{tabular}
\par\smallskip \parbox{0.95\linewidth}{\footnotesize Legend: LMM 1 is correctly specified under a linear covariate effect but intentionally misspecified under nonlinear trends, reflecting settings in which the true functional form is unknown. LMMs 2-4 include fixed categorical time effects~\cite{Hussey2007}, with LMMs 3 and 4 additionally adjusting for baseline or time-updated covariates.}
\end{table}
\FloatBarrier

\subsection{Performance measures}
We performed $n_{sim} = 1,000$ iterations per parameter combination. For each simulated dataset and analysis model, we extracted the estimated fixed intervention effects and other parameters in order to calculate the performance measures defined in Table~\ref{tab:performance_measures}~\cite{Morris2019}.

\begin{table}[!htb]
\caption{Summary of performance measures with their definitions and estimation formulas.}
\label{tab:performance_measures}
\centering
\renewcommand{\arraystretch}{1.3}
\begin{tabular}{p{4cm}p{3.15cm}p{4.5cm}}
    \toprule
    \textbf{Performance measures} &\textbf{Definition} & \textbf{Estimate} \\
    \midrule
    Bias & $E(\hat{\theta}) - \theta$ & $\frac{1}{n_{sim}} \sum_{i=1}^{n_{sim}} \hat{\theta}_i - \theta$ \\
    Average estimated model-based SE & $E\left(\sqrt{\widehat{Var}(\hat{\theta})}\right)$ & $\frac{1}{n_{sim}} \sum_{i=1}^{n_{sim}} \sqrt{\widehat{Var}(\hat{\theta}_i)}$ \\
    Empirical (Monte Carlo) SE & $\sqrt{Var(\hat{\theta})}$ & $\sqrt{\frac{1}{n_{sim}-1} \sum_{i=1}^{n_{sim}} (\hat{\theta}_i-\overline{\theta})^2}$ \\
    Percentage error of average estimated model-based SE to empirical SE & $100\left(\frac{E\left(\sqrt{\widehat{Var}(\hat{\theta})}\right)}{\sqrt{Var(\hat{\theta})}} -1 \right)$ & $100\left(\frac{\frac{1}{n_{sim}} \sum_{i=1}^{n_{sim}} \sqrt{\widehat{Var}(\hat{\theta}_i)}}{\sqrt{\frac{1}{n_{sim}-1} \sum_{i=1}^{n_{sim}} (\hat{\theta}_i - \overline{\theta})^2}} - 1 \right)$ \\
    Coverage & $P(\hat{\theta}_{low} < \theta < \hat{\theta}_{upp})$ & $\frac{1}{n_{sim}} \sum_{i=1}^{n_{sim}} \mathbbm{1}_{\{\hat{\theta}_{low,i} < \theta < \hat{\theta}_{upp,i}\}}$ \\
    Statistical power or type I error & $Pr(p_i \leq \alpha)$ & $\frac{1}{n_{sim}} \sum_{i=1}^{n_{sim}}\mathbbm{1}_{\{p_i \leq \alpha\}}$ \\
    \botrule
\end{tabular}
\par\smallskip \parbox{0.95\linewidth}{\footnotesize Legend: $\hat{\theta}$ represents the estimated parameter, $\theta$ is the true parameter value, $\widehat{Var}(\hat{\theta})$ denotes either the estimated model-based variance or model-robust variance of $\hat{\theta}$, $n_{sim}$ is the number of simulation iterations, $\hat{\theta}_{low,i}$ and $\hat{\theta}_{upp,i}$ denote the lower and upper confidence interval limits, respectively, for simulation $i$, calculated using either the standard or model-robust variance estimator of $\hat{\theta}$, $\mathbbm{1}_{\{\hat{\theta}_{low,i} < \theta < \hat{\theta}_{upp,i}\}}$ is an indicator function that equals 1 if the true parameter value $\theta$ lies within the estimated confidence interval in simulation $i$, and 0 otherwise, $\overline{\theta}$ is the average of the estimated parameters across simulations, and $\mathbbm{1}{\{p_i \leq \alpha\}}$ is an indicator function that equals 1 if the p-value $p_i$ is less than or equal to the significance level $\alpha$, and 0 otherwise. Results for the average estimated model-based SE, empirical (Monte Carlo) SE, and the percentage error of the average estimated model-based SE relative to the empirical SE are presented in the Supplementary Material to improve readability and maintain focus on the primary performance measures in the main manuscript.}
\end{table}
\FloatBarrier

Coverage was assessed as the proportion of simulation runs in which the 95\% confidence interval for the intervention effect contained the true parameter value, thereby evaluating the validity of statistical inference. The proportions of simulations with $p \leq 0.05$ represent the type I error under the null hypothesis ($\theta = 0$) and the statistical power under the alternative hypothesis, respectively. P-values were obtained from a Wald test using a Satterthwaite's degrees of freedom (DF) approximation, which provides more accurate DFs for LMMs with complex variance structures and improves small-sample inference~\cite{Pustejovsky2018, Hemming2025_2}. Furthermore, we visualized the distribution of the estimated fixed intervention effects using boxplots.

\subsection{Cluster-robust variance estimators}
In addition to standard variance estimators, we used cluster-robust variance estimators (CRVE). In LMMs, CRVEs improve the robustness of standard error estimates of the intervention effect when the linear predictor is misspecified or when assumptions about the likelihood are violated, such as in the presence of heteroscedasticity or residual dependence. While LMM fixed-effect estimates are generally robust to certain forms of misspecification, standard inference may not be. This makes CRVEs a useful approach for obtaining more reliable standard errors and confidence intervals.
\par
CRVEs offer a flexible approach to estimating the covariance matrix, ensuring validity even when the likelihood is misspecified. CRVEs are also known as \enquote{sandwich} estimators due to their mathematical form, consisting of two outer \enquote{bread} matrices (derived from the Hessian of the log-likelihood) and an inner \enquote{meat} matrix (based on cross-products of the corresponding score function). For a comprehensive overview of cluster-robust inference, see Cameron and Miller (2015)~\cite{Cameron2015} and Pustejovsky and Tipton (2018)~\cite{Pustejovsky2018}.
\par
However, standard CRVE approaches may perform poorly when the number of clusters is small (e.g., fewer than ten), which often leads to downward-biased standard errors and inflated type I error rates~\cite{Cameron2015}. To mitigate this issue, several small-sample corrections have been proposed. In particular, CR2VE and CR3VE incorporate bias-reducing adjustments to improve finite-sample performance. CR2VE applies an analytic correction to reduce small-sample bias in the estimated covariance matrix~\cite{Bell2002}. In contrast, CR3VE uses a leave-one-cluster-out approximation, which is analogous to the jackknife, and tends to provide more conservative inference. Pustejovsky and Tipton (2018) demonstrated that, especially when combined with Satterthwaite degrees of freedom corrections, CR2VE can substantially improve the performance of hypothesis testing in finite samples. However, accuracy still depends on the number of clusters and the characteristics of the study design~\cite{Pustejovsky2018}. CR3VE may further reduce type I error inflation in very small samples, albeit sometimes at the cost of reduced statistical power~\cite{Pustejovsky2018}.
\par
In this simulation, we use three types of CRVEs: CR0VE, CR2VE, and CR3VE. These estimators differ primarily in the construction of the adjustment applied to the \enquote{meat} matrix and were implemented in the R package \texttt{clubSandwich}~\cite{clubSandwich}.

\section{Results}\label{sec:results}
\subsection{Bias and distribution of intervention effect estimates}
\begin{figure*}[!htb]
    \centering
    \begin{subfigure}[b]{0.9\textwidth}
        \centering\includegraphics[width=\textwidth]{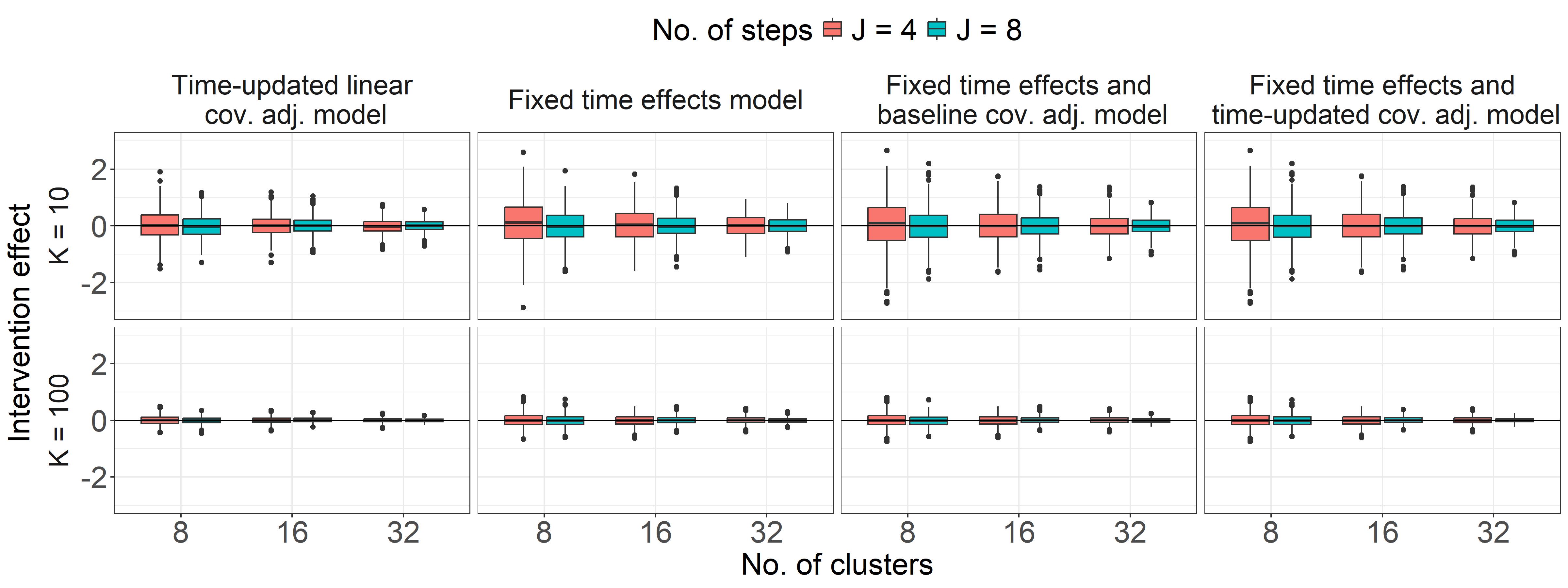}
        \caption{Linear influence}
        \label{subfig:Estimate_closedCohort_linear_converged}
    \end{subfigure}
    \vspace{0.5cm} % Adjust the space between subfigures if needed
    \begin{subfigure}[b]{0.9\textwidth}
        \centering\includegraphics[width=\textwidth]{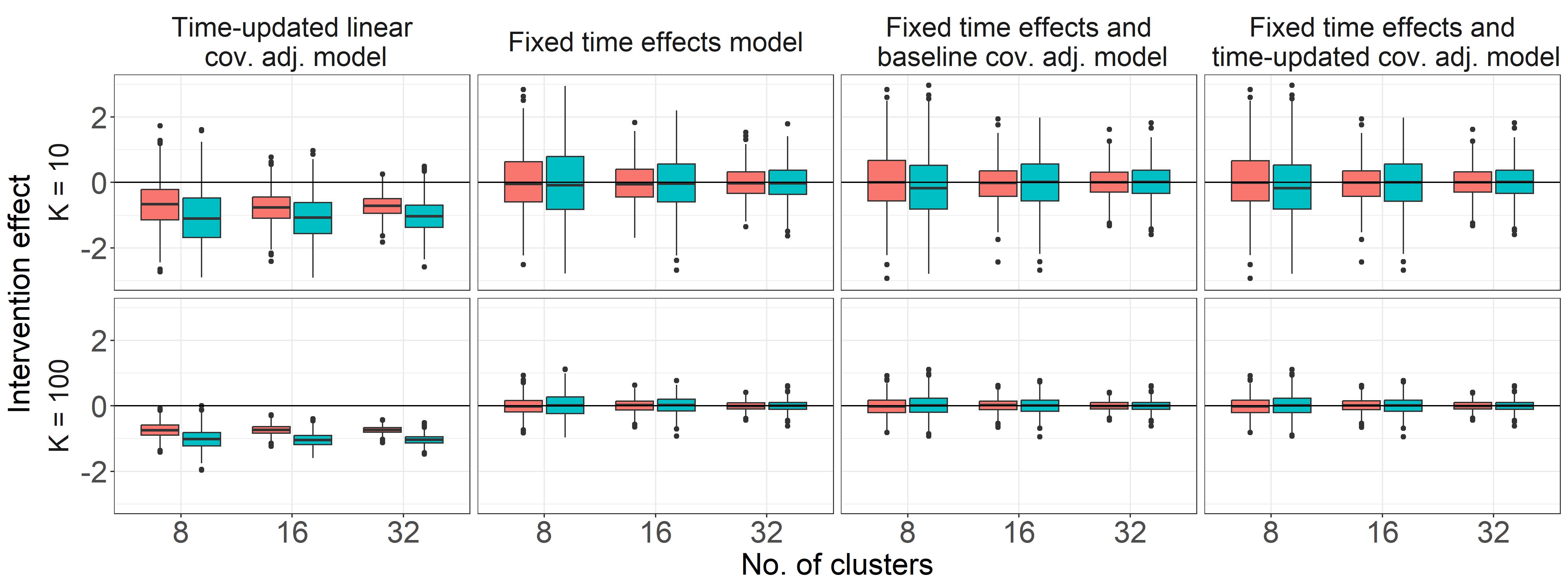}
        \caption{Nonlinear influence}
        \label{subfig:Estimate_closedCohort_nonlinear_converged}
    \end{subfigure}
    \caption{Estimated intervention effects from 1,000 simulations per parameter combination ($I= 8, 16, 32$; $K = 10, 100$; $J = 4, 8$) using four analysis models (Table~\ref{tab:mixed_models}; cov. adj. = covariate adjustment), including only converged models. Closed cohort data with a covariate affecting the outcome (a) linearly and (b) nonlinearly.}
    \label{fig:Estimate_closedCohort}
\end{figure*}

Figure~\ref{fig:Estimate_closedCohort} shows the distribution of up to 1,000 estimates of $\theta$ obtained from converged models only.
\par
When the time-dependent covariate effect is linear, the estimated intervention effects are symmetrically distributed around the true value of zero ($\theta = 0$) and hence approximately unbiased (Supplementary Table~\ref{tab:Bias}) for each model (Figure~\ref{subfig:Estimate_closedCohort_linear_converged}). However, with a nonlinear covariate effect, the model without fixed categorical time effects (Equation 1) underestimates the true intervention effect (Figure~\ref{subfig:Estimate_closedCohort_nonlinear_converged}). The other three models, which include fixed categorical time effects, exhibit estimated intervention effects that are also symmetrically distributed around the true value and hence unbiased (Supplementary Table~\ref{tab:Bias}).
\par
Models with fixed categorical time effects (Equations 2-4) have nearly identical distributions of $\hat{\theta}$, regardless of whether or how covariate adjustments are made.
\par
Increasing the sample size, in terms of both the number of clusters ($I$) and individuals per cluster-period ($K$), reduces the interquartile range and the overall variability of the estimates. Extending the observation period from four to eight steps ($J$), thereby raising the total number of observations, improves the precision of intervention effect estimates when the covariate has a linear effect on the outcome. However, this advantage does not apply when the covariate affects the outcome nonlinearly.

\FloatBarrier

In our simulations, we excluded results when models did not converge. Supplementary Figure~\ref{fig:Estimate_closedCohort_withnonconverged} presents estimates from both converged and non-converged models and suggests that no systematic differences were apparent in the distribution of intervention effect estimates between these cases. The distribution of the estimated intervention effect is quite similar for the open cohort (Supplementary Figure~\ref{fig:Estimate_openCohort}).

\subsection{Coverage and type I error}
Since the model in Equation 2 provides unbiased estimates of the intervention effect and additional covariate adjustments do not reduce SEs, we will focus on this model for statistical inference. Coverage, percentage standard error bias, type I error rate and statistical power of models specified in Equation 1, 3, and 4 are shown in Supplementary Tables~\ref{tab:Coverage_linear} to~\ref{tab:Power_nonlinear}. All results in this section are based on simulations under the null hypothesis ($\theta = 0$), allowing type I error rates to be directly assessed.
\par
\begin{figure}[!htb]
    \centering
    \begin{subfigure}[b]{0.49\textwidth}
        \includegraphics[width=\textwidth]{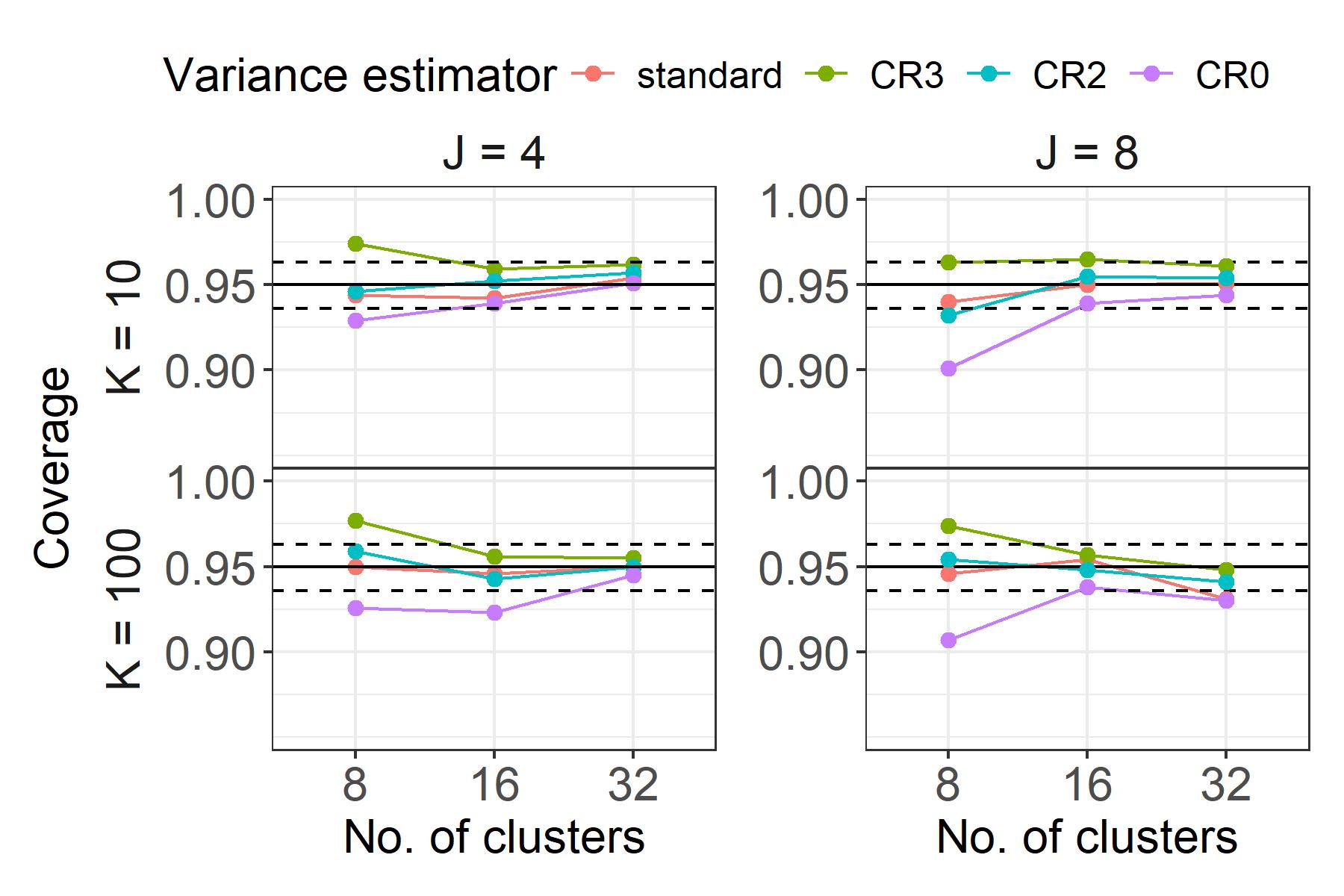}
        \caption{Closed cohort and linear influence}
        \label{fig:Coverage_closedCohort_linear}
    \end{subfigure}
    \hfill
    \begin{subfigure}[b]{0.49\textwidth}
        \includegraphics[width=\textwidth]{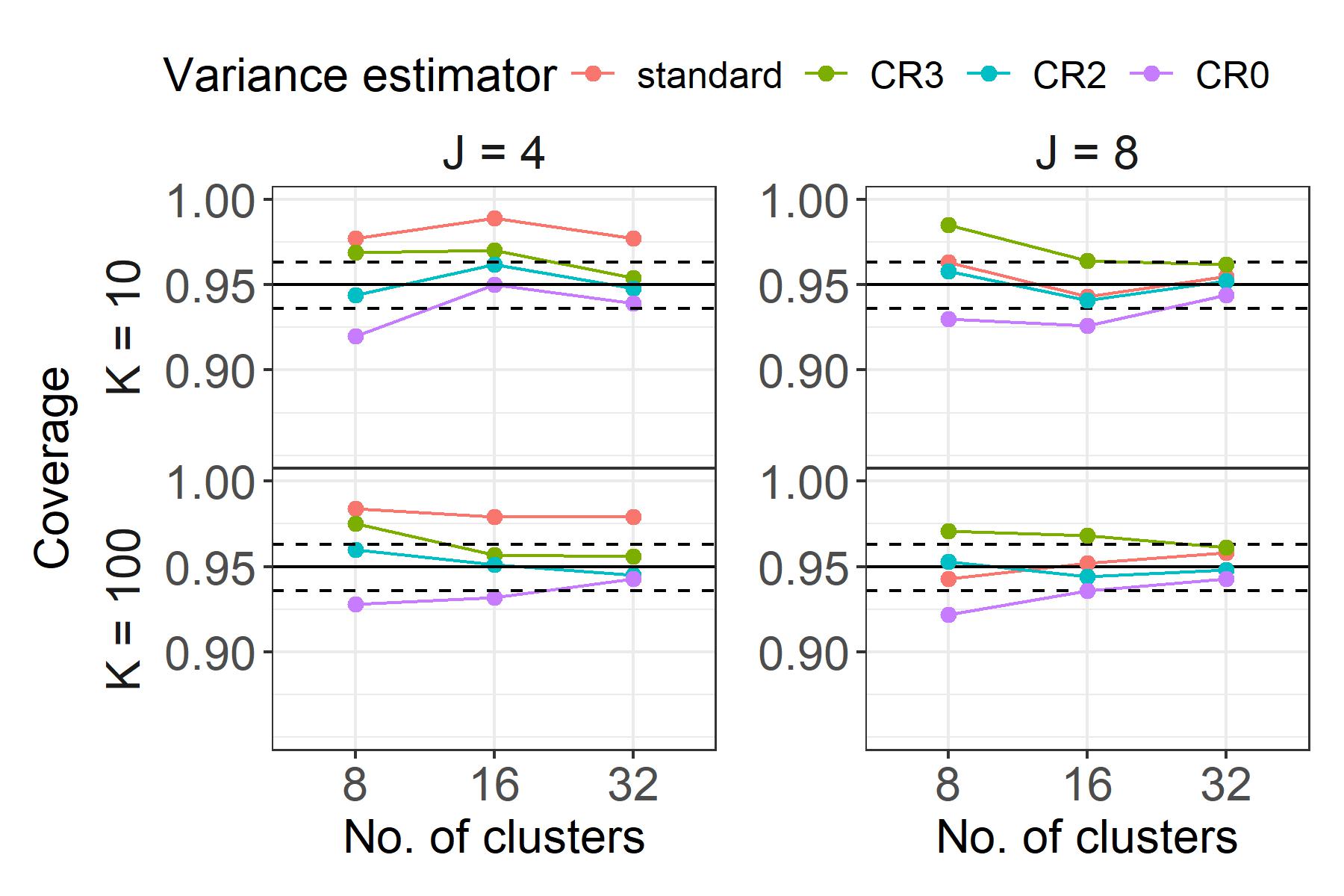}
        \caption{Closed cohort and nonlinear influence}
        \label{fig:Coverage_closedCohort_nonlinear}
    \end{subfigure}
    \centering
    \begin{subfigure}[b]{0.49\textwidth}
        \includegraphics[width=\textwidth]{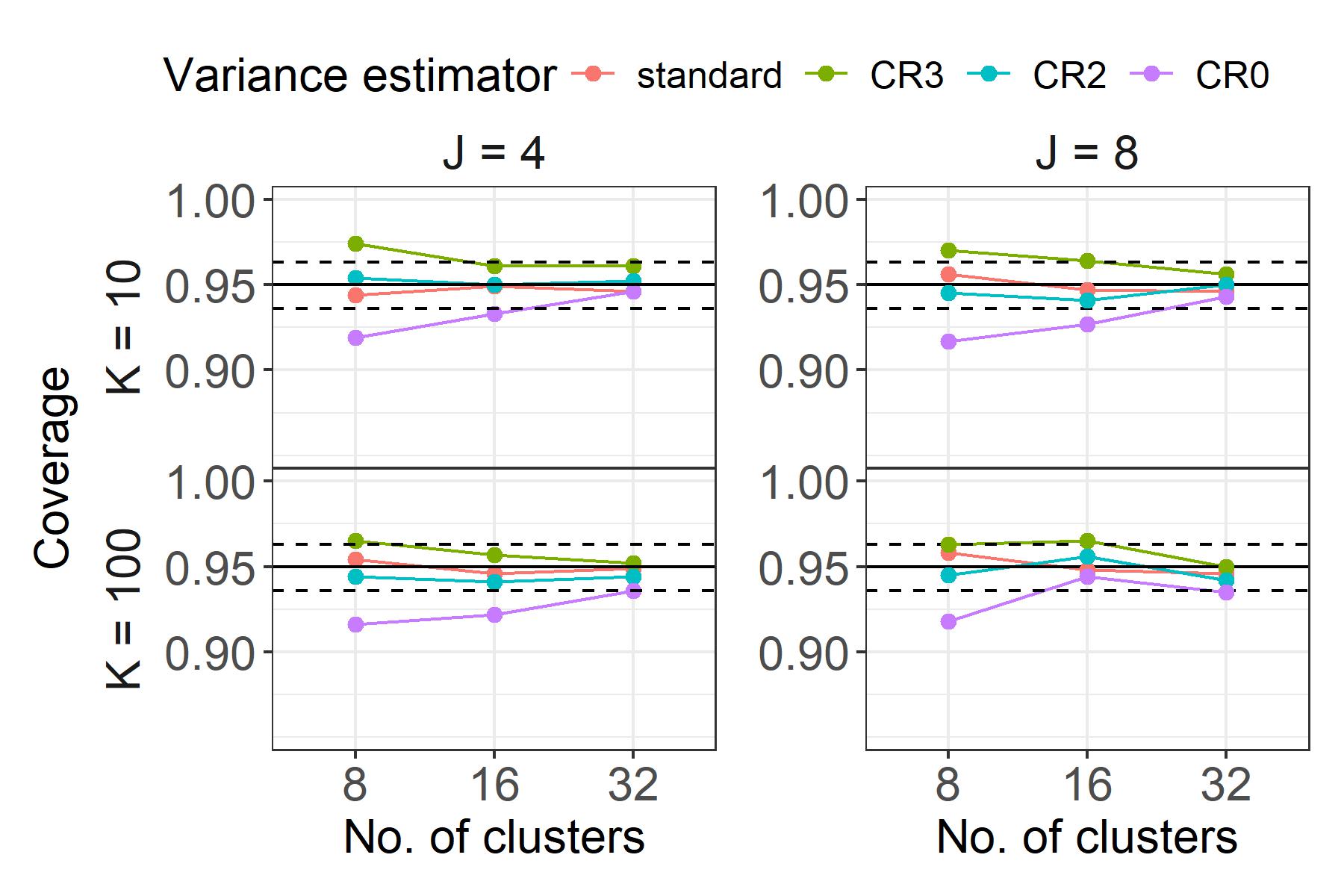}
        \caption{Open cohort and linear influence}
        \label{fig:Coverage_openCohort_linear}
    \end{subfigure}
    \hfill
    \begin{subfigure}[b]{0.49\textwidth}
        \includegraphics[width=\textwidth]{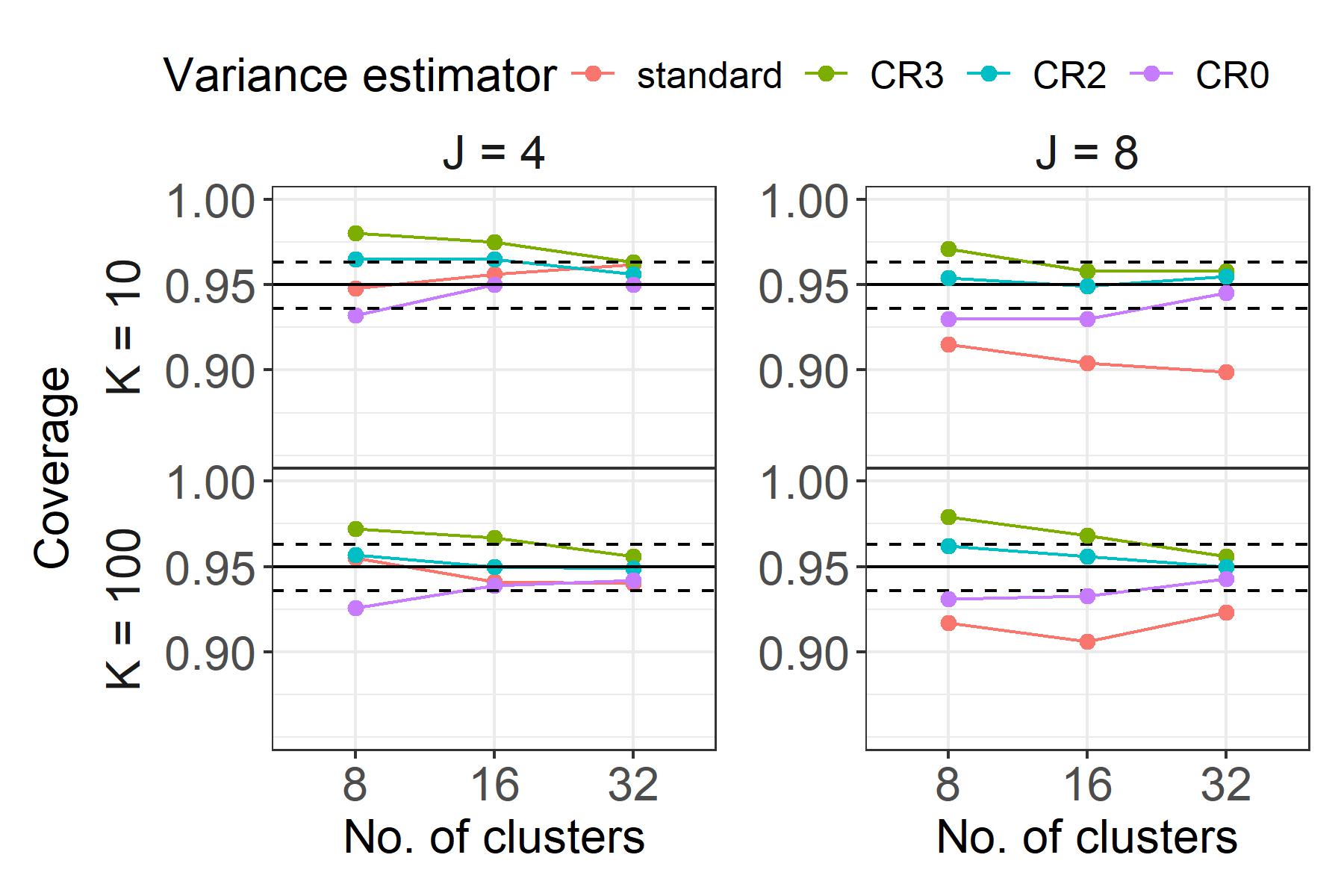}
        \caption{Open cohort and nonlinear influence}
        \label{fig:Coverage_openCohort_nonlinear}
    \end{subfigure}
    \caption{Coverage from 1,000 simulations per parameter combination ($I = 8, 16, 32; K = 10, 100; J = 4, 8$). The analysis model includes fixed categorical time effects (Equation 2) and is presented with standard estimation (red line) and three cluster-robust variance estimation methods: CR0VE (purple line), CR2VE (blue line), and CR3VE (green line). Results are displayed for four scenarios: (a) closed cohort with linear covariate influence, (b) closed cohort with nonlinear covariate influence, (c) open cohort with linear covariate influence, and (d) open cohort with nonlinear covariate influence. The solid line indicates the nominal 95\% coverage level, while the dashed lines show Monte Carlo uncertainty, approximated as 95\% $\pm$ 2 MCSE under a binomial model with 1,000 simulation.}
    \label{fig:Coverage_all}
\end{figure}

Figure~\ref{fig:Coverage_all} shows that models employing standard variance estimation (without cluster-robust variance estimation; red line) achieve coverage close to the nominal 95\% level when the time-varying covariate has a linear effect on the outcome, both in closed and open cohort settings (Subfigs.~\ref{fig:Coverage_closedCohort_linear} and~\ref{fig:Coverage_openCohort_linear}, respectively). However, coverage deviates from the nominal level when the covariate effect is nonlinear (Subfigs.~\ref{fig:Coverage_closedCohort_nonlinear} and~\ref{fig:Coverage_openCohort_nonlinear}), resulting in either conservative coverage (above 95\%) or undercoverage (below 95\%), depending on the parameter combination.
\par
Across all parameter combinations and settings in our simulation, the model using CR0 variance estimation (purple line) exhibits substantial undercoverage, particularly in scenarios with a small number of clusters. In contrast, models using the CR3VE (green line) tend to yield slightly conservative coverage, especially when the number of clusters is small. Models with CR2 variance estimation (blue line) generally achieve coverage closest to the nominal level, although some deviations persist for certain parameter combinations.

\FloatBarrier

\begin{figure}[!htb]
    \centering
    \begin{subfigure}[b]{0.49\textwidth}
        \includegraphics[width=\textwidth]{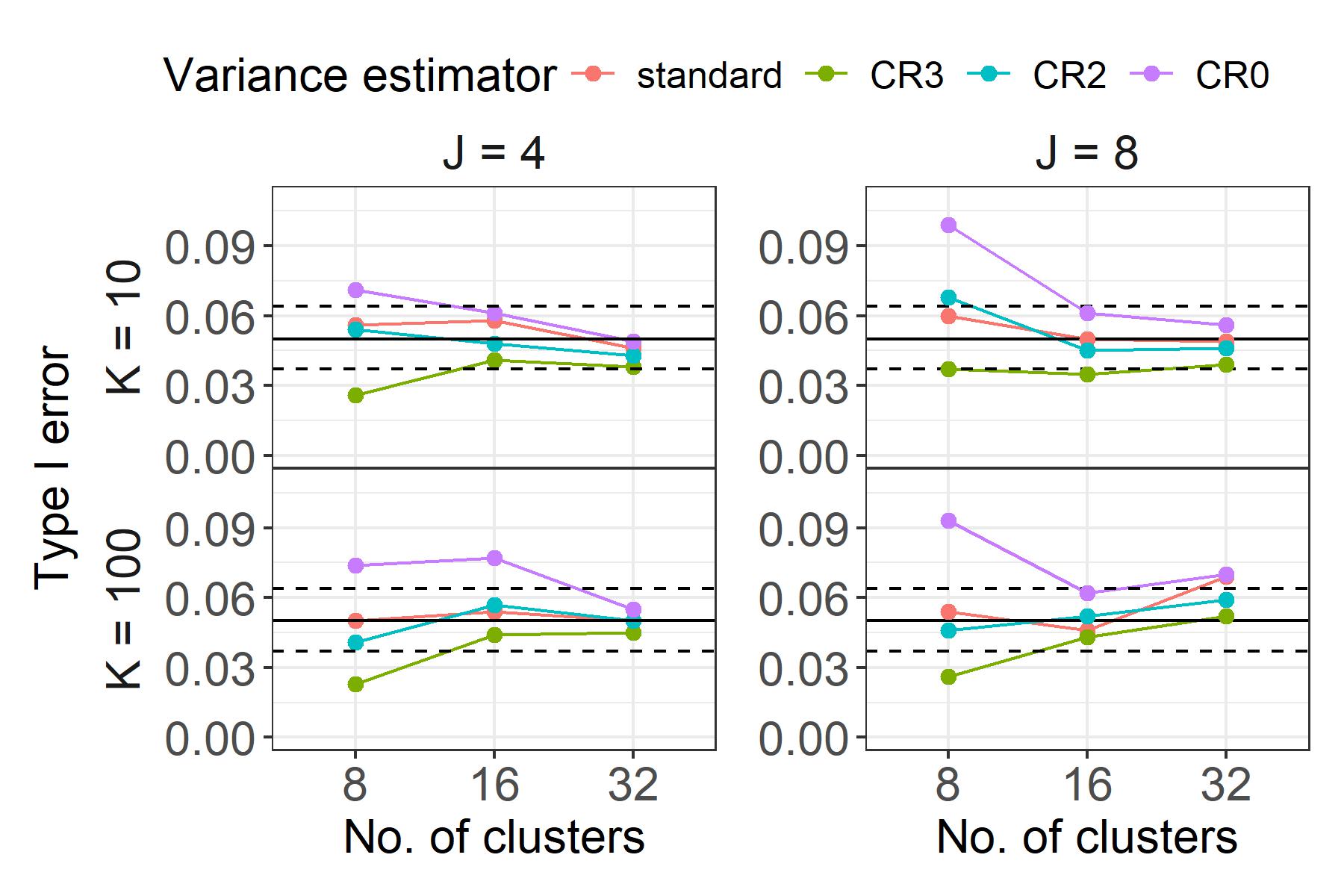}
        \caption{Closed cohort and linear influence}
        \label{fig:TypeIError_closedCohort_linear}
    \end{subfigure}
    \hfill
    \begin{subfigure}[b]{0.49\textwidth}
        \includegraphics[width=\textwidth]{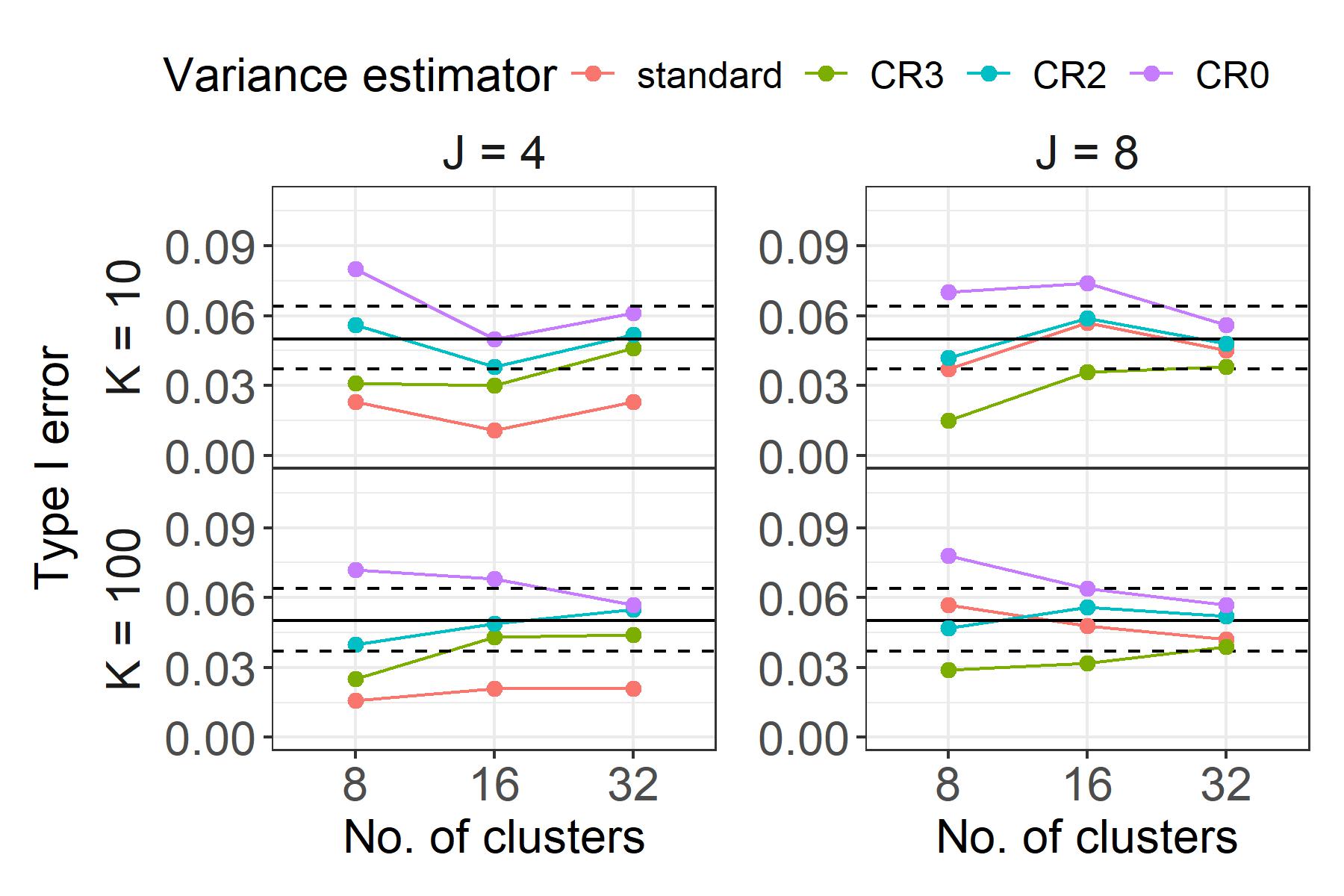}
        \caption{Closed cohort and nonlinear influence}
        \label{fig:TypeIError_closedCohort_nonlinear}
    \end{subfigure}
    \centering
    \begin{subfigure}[b]{0.49\textwidth}
        \includegraphics[width=\textwidth]{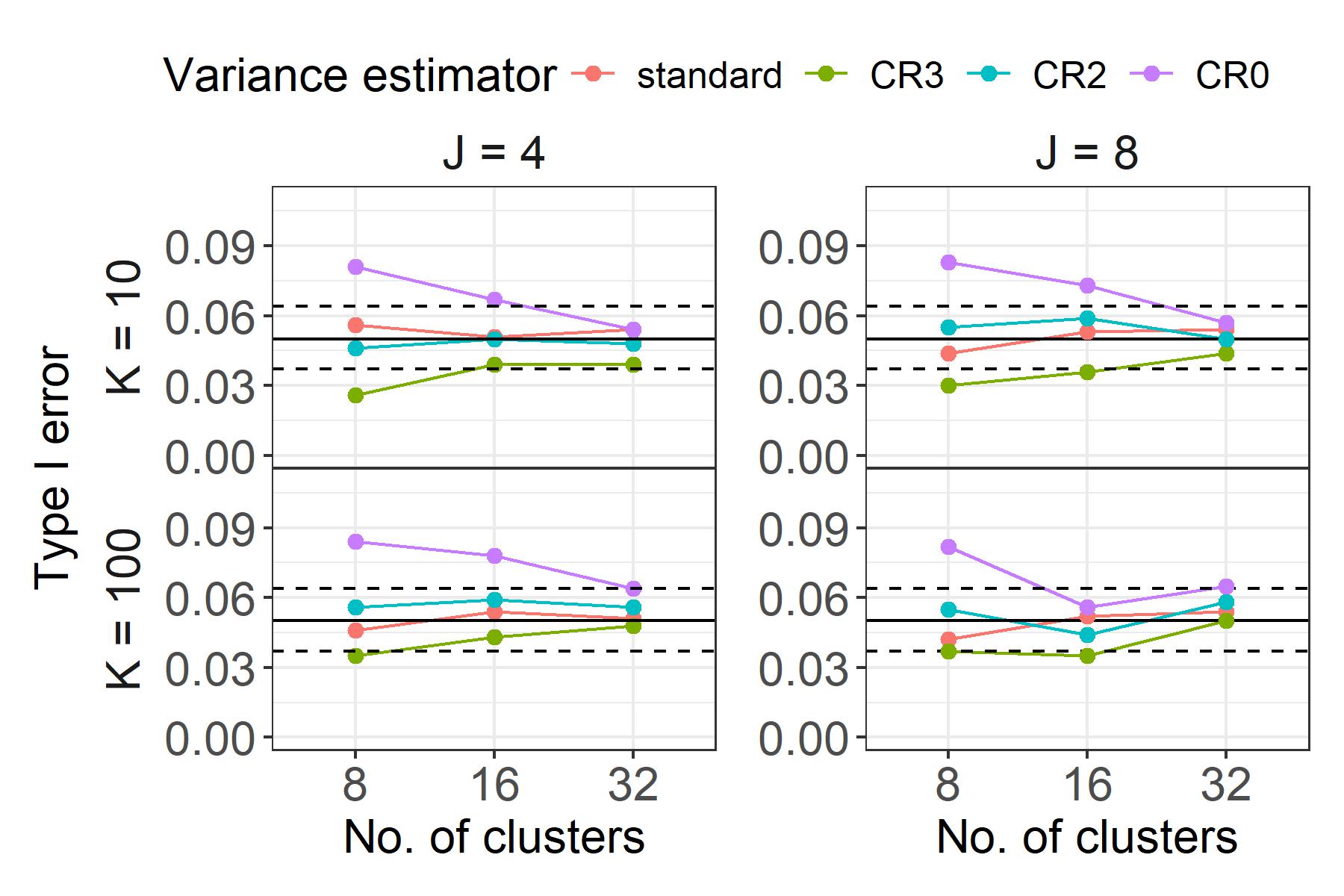}
        \caption{Open cohort and linear influence}
        \label{fig:TypeIError_openCohort_linear}
    \end{subfigure}
    \hfill
    \begin{subfigure}[b]{0.49\textwidth}
        \includegraphics[width=\textwidth]{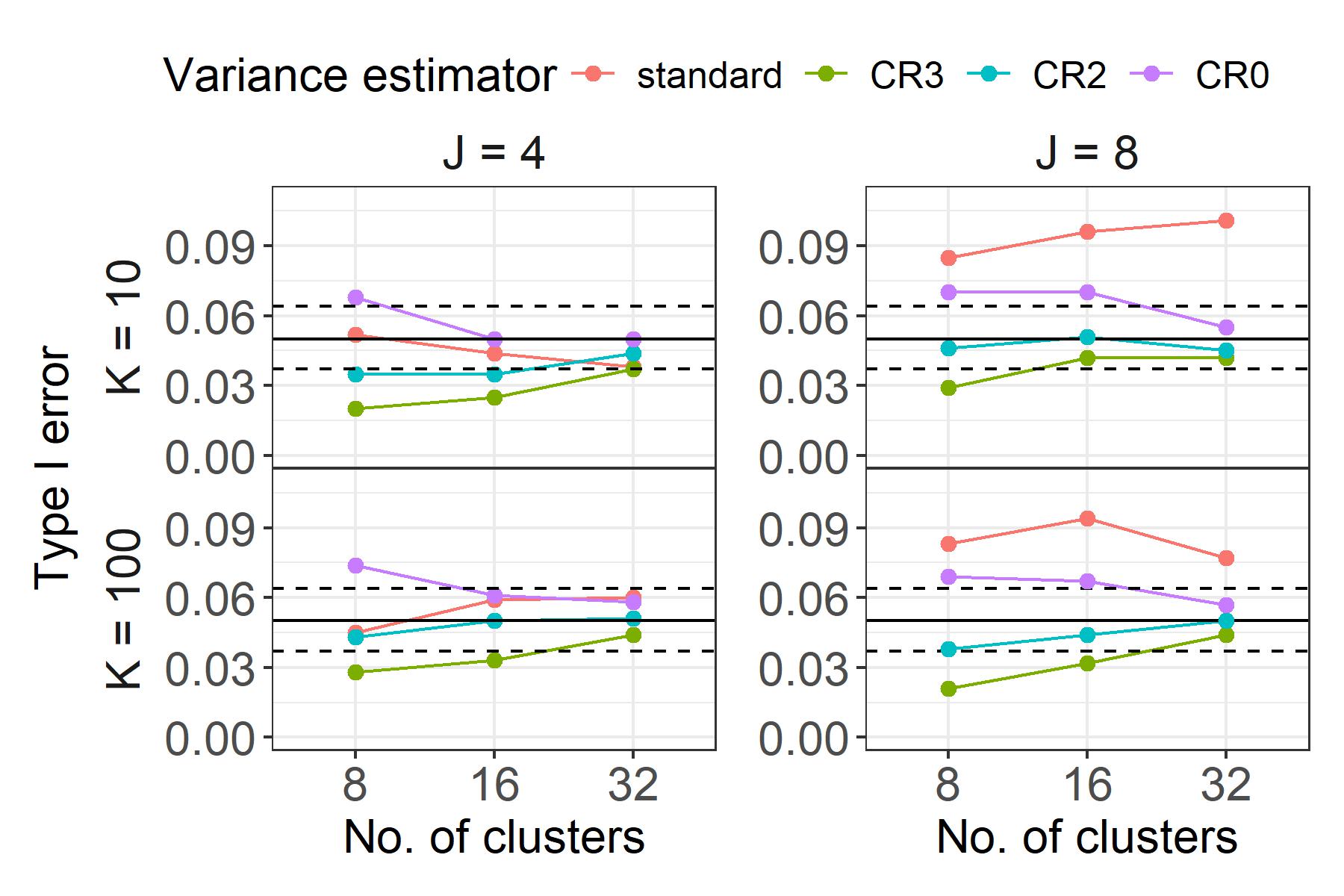}
        \caption{Open cohort and nonlinear influence}
        \label{fig:TypeIError_openCohort_nonlinear}
    \end{subfigure}
    \caption{Type I error from 1,000 simulations per parameter combination ($I = 8, 16, 32; K = 10, 100; J = 4, 8$). The analysis model includes fixed categorical time effects (Equation 2) and is presented with standard estimation (red line) and three cluster-robust variance estimation methods: CR0VE (purple line), CR2VE (blue line), and CR3VE (green line). Results are displayed for four scenarios: (a) closed cohort with linear covariate influence, (b) closed cohort with nonlinear covariate influence, (c) open cohort with linear covariate influence, and (d) open cohort with nonlinear covariate influence.}
    \label{fig:TypeIError_all}
\end{figure}

Undercoverage of confidence intervals corresponds to inflated type I error rates under the null hypothesis ($\theta = 0$), while overcoverage results in more conservative inference and reduced statistical power. Although the latter is undesirable, overcoverage is generally preferable to undercoverage, as it better preserves type I error control. Figure~\ref{fig:TypeIError_all} shows how deviations from nominal coverage translate into the empirical type I error rate. The solid horizontal line at $\alpha = 0.05$ denotes the nominal significance level, i.e. the probability of incorrectly rejecting the null hypothesis when it is true. The dashed lines show the 2.5th and 97.5th percentiles of the binomial distribution with parameters $n = 1,000$ and $p = 0.05$, representing the central 95\% range of expected rejections across 1,000 simulations under the null hypothesis.
\par
In the case of a nonlinear individual time effect, the standard VE (without cluster-robust variance estimation; red line) showed overly conservative type I error control at four steps and failed to maintain the type I error rate at eight steps. Even in the linear case, the CR0VE (purple line) does not control the type I error rate adequately. The CR2VE (blue line) maintains type I error for most, but not all, parameter settings. In contrast, the CR3VE (green line) controls type I error consistently across all parameter settings. However, it is slightly conservative, especially when the number of clusters is small.

\FloatBarrier

\subsection{Statistical power and efficiency}
The three analysis models that incorporate fixed categorical time effects (Equation 2-4) have comparable statistical power. However, their statistical power is lower than that of the data-generating model (Equation 1) when the covariate has a linear effect on the outcome (Supplementary Figures~\ref{fig:Power_closedCohort_linear} and~\ref{fig:Power_closedCohort_nonlinear}). This reflects lower efficiency when a correctly specified parametric covariate model is replaced by more flexible fixed categorical time effects, which are needed to adequately account for calendar and individual time effects.

\subsection{Convergence and singular fit behavior}
Supplementary Tables~\ref{tab:SingFit_NonConv_Eq1} - \ref{tab:SingFit_NonConv_Eq4} summarize the proportions of singular fits and non-convergence across analysis models and parameter settings (\texttt{bobyqa} optimizer). Both issues occurred across all models, with the highest frequencies observed when the number of clusters was small ($I=8$). For the model including fixed categorical time effects (Equation 2), non-convergence rates were consistently high. When $I=8$ and $K=10$, approximately 47-52\% of models failed to converge, with similar rates observed for both linear and nonlinear covariate effects. For larger cluster-period sizes ($K=100$), non-convergence remained substantial, ranging from approximately 29-40\%.
\par
For models without fixed categorical time effects (Equation 1) and those including additional covariate adjustment (Equations 3 and 4), convergence behavior depended strongly on the functional form of the covariate. Under linear effects, singular fits and non-convergence were rare for $K=10$ (generally $\textless$1\%), but non-convergence increased substantially for $K=100$, reaching up to approximately 60–67\%. Under nonlinear effects, both singular fits (up to 36\%) and non-convergence (up to 47\%) were frequent for smaller cluster size, while singular fits became less common but non-convergence remained prevalent when $K=100$.

\section{Discussion}\label{sec:discussion}
To date, the methodological focus of SW-CRTs has primarily been on cross-sectional designs, with limited guidance available for the evaluation of SW-CRTs using cohort designs~\cite{Barker2016}. While several simulation studies~\cite{Rennert2021, Nickless2018} have examined the handling of calendar time effects in cross-sectional SW-CRTs, cohort designs introduce an additional temporal dimension, namely individual time effects, that must be addressed. As the motivating example in Section~\ref{sec:motivation} demonstrates, failure to model these individual time effects can, in certain settings, bias estimation of the intervention effect.
\par
To the best of our knowledge, our simulation study is the first to systematically investigate the influence of individual time effects in cohort SW-CRTs. Across the scenarios considered, LMMs specifying a fixed intervention effect, fixed categorical time effects, and two random intercepts for clustering and repeated measurements of individual-level outcomes (Equation 2) yielded unbiased intervention effect estimates in the scenarios considered when the random effects structure aligns with the model that generated the data. This finding holds true for both closed and open cohort data, even when individual-level changes over time were unmeasured or their functional form was unknown. This is a major advantage because it reduces the need to explicitly model individual time effects. However, the absence of fixed categorical time effects (Equation 1) may result in biased estimates if the covariate is not modeled according to its functional influence.
\par
Equation 2 yielded unbiased intervention effect estimates in our simulations, even without further covariate adjustment, as fixed categorical time effects capture the temporal changes in the cohort, and the random individual effects account for differences in participants' baseline characteristics. However, linear and nonlinear time-varying covariates differently affect statistical inference. Linear time-varying covariates have a constant effect over time and across all individuals, regardless of their baseline values. The fixed categorical time effects thus capture the change in characteristics over time equally for all individuals. In contrast, nonlinear time-varying covariates lead to heteroscedasticity of residuals, as categorically modeled time effects depict the changes with varying degrees of precision depending on the participants' baseline characteristics (e.g., difference in age at the start of the study). Therefore, the Gauss Markov assumption A4 $\epsilon_{ijk} \sim i.i.d(0,\sigma_e^2)$ is violated, which requires each residual to exhibit constant variance independent of the covariates of the LMM. When time-varying covariates have a nonlinear effect, standard variance estimation can lead to biased inference, reflected in deviations from nominal coverage and corresponding miscontrol of the type I error rate. In such settings, cluster-robust variance estimation appeared necessary to achieve nominal coverage. In contrast, when the effect of time-varying covariates is linear, standard variance estimation achieves coverage close to the nominal level and maintains the type I error rate. Among the various types of CRVEs, the CR3VE (also called Mancl \& DeRouen VE) combined with Satterthwaite's DF correction consistently yielded coverage close to the nominal level and controlled the type I error rate across the scenarios considered. However, it tends to be slightly conservative, particularly when the number of clusters is small, which is reflected in mild overcoverage and a corresponding loss of statistical power. This illustrates the general trade-off of CRVE methods between improved type I error control and reduced efficiency in small samples.
\par
Convergence problems and singular fits primarily occurred in scenarios with a small number of clusters and weakly identified variance components. In these settings, cluster-level variances were often estimated as close to zero, resulting in boundary estimates and numerical instability. All models were fitted using the  \texttt{bobyqa} optimizer (the default in  \texttt{lmerTest}), and alternative optimizers (e.g., \texttt{Nelder-Mead}) did not meaningfully improve convergence. This suggests that the problems are structural rather than purely numerical. One likely reason is unmodeled heterogeneity at the individual level. Without covariate adjustment, baseline differences between individuals can lead to substantial variation in individual random intercepts while cluster-level variance remains small and is often estimated as zero. This imbalance can result in singular fits, which frequently occur alongside convergence problems. Even when singular fits do not occur, convergence issues may arise in settings with little information or when variance components are weakly supported by the data. Small changes in the optimization settings (e.g., increasing iteration limits) may help in some cases, but they are unlikely to resolve problems arising from weak identifiability. Therefore, convergence warnings should be considered together with singular fit indicators, as both typically point to overly complex or weakly supported random effects structures. Residual diagnostics can help detect important covariates that are not yet included in the model. Including such covariates can reduce individual-level variability and improve the estimation of cluster-level variance components.
\par
To evaluate the impact of convergence issues, we compared intervention effect estimates from models that converged with those that did not and observed no meaningful differences in their mean or variability. However, this finding is conditional on the observed outputs and should not be interpreted as implying that convergence issues can be disregarded in applied analyses.
\par
Due to the complexity of cohort SW-CRTs, our simulation study required several assumptions to ensure tractability and interpretability.
\par
First, in a closed cohort design, we assumed equal cluster sizes. However, cluster size can be informative which means that outcomes and/or treatment effects differ according to the cluster size. Previous simulation studies by Martin et al. (2019)~\cite{Martin2019} and Ouyang et al. (2020)~\cite{Ouyang2020} have shown that unequal cluster sizes can affect the precision of intervention effects estimates and, consequently, the statistical power of SW-CRTs in cross-sectional designs. It is reasonable to posit that this finding may also apply to cohort SW-CRTs. However, due to the recruitment of participants at baseline in closed cohorts, clusters are more likely to exhibit consistent sizes than in cross-sectional studies, in which new participants are enrolled at each period.
\par
Second, consistent with the outcomes examined in our motivational example~\cite{Muntinga2012, Hoogendijk2015} and other simulation studies~\cite{Nickless2018, Ouyang2020, Grantham2022}, our simulation study focused on continuous outcomes. In research areas such as health services research, where SW-CRTs are commonly employed~\cite{Matthews2017}, patient-reported outcome measures are increasingly being used to compare intervention groups~\cite{Kohlmann2010}. Typically, differences in group means are assessed using sum-score-based approaches, which are predominantly continuous measures. However, it is important to note that violations of assumptions regarding the distribution of random effects can have minor consequences for LMMs with continuous outcomes, but can have severe implications for generalized linear mixed models (GLMMs)~\cite{Litiere2007, Grilli2015, ArangoBotero2022, Vu2025}. Therefore, our finding may not directly generalize to settings with binary or count outcomes, and additional caution is warranted when extending these results to such contexts.
\par
Third, our modeling approach used a random intercept model which assumes constant time and intervention effects across all clusters. However, trialists may need to consider more complex random effects structures. For instance, they could include a random effect for the time period~\cite{Thompson2017_B} or adopt a random cluster-by-period model~\cite{Ouyang2024}. In their simulation study, Ouyang et al. (2024)~\cite{Ouyang2024} demonstrated that using a robust variance estimator with either the random intercept or random cluster-by-period models can yield valid statistical inference for fixed effects, even when the random effects structure is misspecified. However, incorporating individual-level random slopes for categorical time effects is often not feasible in SW-CRTs, as it can lead to an overparameterized and non-identifiable model when the number of random effects approaches the number of observations.
\par
Our simulation study corroborates the findings of Schielzeth et al. (2020)~\cite{Schielzeth2020} and Wang et al. (2024)~\cite{Wang2024b} that LMMs are remarkably robust and valid inference is obtained via the cluster-robust variance estimation.

\section{Conclusions and recommendations}\label{sec:conclusion}
This article highlights the influence of individual time effects in cohort SW-CRTs. Despite the fact that the CONSORT guideline acknowledges time effects as an important potential confounder~\cite{Hemming2019}, this aspect has received limited attention. To improve understanding and application of SW-CRTs, we recommend that the CONSORT guideline explicitly differentiate between cross-sectional and cohort designs. This distinction is important because both designs are susceptible to secular trends (calendar time effects) and cluster-level changes, but cohort designs are also subject to individual time effects. These individual time effects can impact the statistical inference of the fixed intervention effect and thereby the validity of trial conclusions.
\par
\begin{figure}[!htb]
    \centerline{\includegraphics[width=\textwidth]{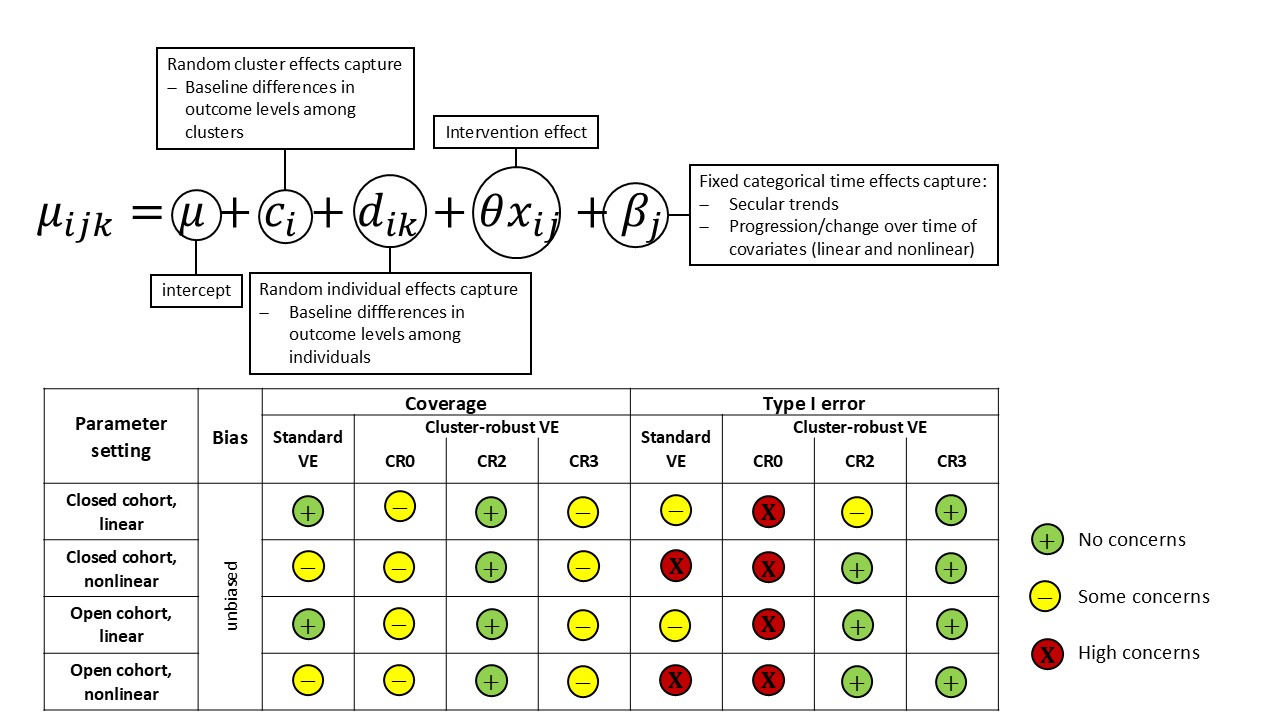}}
    \caption{Main simulation results. Equation 2 yielded unbiased intervention effect estimates for both closed and open cohort data, even with unmeasured or misspecified individual time effects. Random individual intercepts absorb baseline heterogeneity, and fixed categorical time effects capture changes over time (e.g., aging). Misspecified residual correlation structures did not bias intervention effect estimates but substantially affected inference for fixed effects. The figure summarizes coverage and type I error rates for models using the standard variance estimator (standard VE) and cluster-robust variance estimators (CR0VE: standard CRVE, CR3VE: bias-reduced correction, CR2VE: bias-reduced linearization method). The CR3VE combined with Satterthwaite's degrees of freedom correction produced variance estimates that resulted in appropriate coverage and type I error control across all parameter settings. Concern levels were assigned qualitatively. \enquote{Some concerns} indicates noticeable under- or overcoverage despite largely acceptable type I error control, whereas \enquote{high concerns} indicates substantial inflation of type I error rates.\label{fig:ResultsOverview}}
\end{figure}

Our findings demonstrate that a LMM including fixed categorical time effects and random intercepts for clusters and individuals provides unbiased intervention effect estimates across a broad range of scenarios. However, valid statistical inference depends on the nature of time-varying covariate effects. When these effects are nonlinear or unknown, as is often the case in practice, standard variance estimation may lead to invalid inference, whereas cluster-robust variance estimation (CR2VE and CR3VE) generally improves type I error control. Among the estimators considered, CR3VE provided the most consistently reliable type I error control across all scenarios and is therefore recommended when strict control of false-positive findings is a priority (Fig.~\ref{fig:ResultsOverview}). However, this increased robustness was accompanied by slightly conservative inference and corresponding reductions in statistical power, particularly in settings with few clusters. In contrast, CR2VE achieved coverage and type I error rates close to nominal levels in most scenarios while generally retaining greater power, making it a reasonable alternative when a less conservative approach is acceptable.

\FloatBarrier

Given that the functional form of time-varying covariates is often unknown in practice, these findings support the routine use of cluster-robust variance estimation. Diagnostic tools such as residual plots may help detect model misspecification, but their ability to distinguish between omitted nonlinear covariates and misspecified correlation structures is limited.
\par
In summary, we recommend fitting models with fixed categorical time effects and random intercepts for clusters and individuals, and using CR3VE for statistical inference, particularly in settings where model misspecification cannot be ruled out. These strategies provide a practical and robust framework for the analysis of cohort SW-CRTs and help ensure valid and reliable conclusions in applied research.

\backmatter

\bmhead{Abbreviations}
ACT: \enquote{frail older Adults: Care in Transition}; CONSORT: Consolidated standards of reporting trials; Cov. adj.: Covariate adjustment; CRT: Cluster randomized trial;  CRVE: Cluster-robust variance estimator; CR0VE: Standard cluster-robust variance estimator; CR2VE: Cluster-robust variance estimator with bias-reduced linearization method; CR3VE: Cluster-robust variance estimator which approximates the leave-one-cluster-out jackknife variance estimator; Eq.: Equation; DF: Degrees of freedom; ICC: Intracluster correlation coefficient; i.i.d. = independent and identically distributed; LMM: Linear mixed model;  MCS: Mental health component score; MCSE: Monte Carlo standard error; PCS: Physical health component score; REML: Restricted maximum likelihood;  SE: Standard error; SW-CRT: Stepped wedge cluster randomized trial; VE: Variance estimator

\bmhead{Acknowledgements}
We would like to thank Tobias Schipp for his contributions to the simulation code. This research was conducted as part of the PhD thesis of Jale Basten at TU Dortmund University.

\bmhead{Authors' contributions}
JB and NT conceived the methods and JB drafted the manuscript. KI and NT critically reviewed the contents of the paper. JB carried out the simulation study. All authors have read and approve the final manuscript.

\bmhead{Funding}
Not applicable.

\bmhead{Availability of data and materials}
All data generated and simulation codes supporting the conclusions of this article are available from the corresponding author on reasonable request.

\section*{Declarations}
\bmhead{Ethics approval and consent to participate}
No ethical approval or consent was required for this simulation study based on a proposed trial design.

\bmhead{Consent for publication}
Not applicable.

\bmhead{Competing interests}
The authors declare that they have no competing interests.

\bibliography{Literatur} % common bib file

\newpage

\section*{Supplementary information}\label{sec:supplement}
\renewcommand{\thetable}{S\arabic{table}}
\renewcommand{\thefigure}{S\arabic{figure}}
\setcounter{table}{0}
\setcounter{figure}{0}
\renewcommand{\thesection}{S\arabic{section}}
\setcounter{section}{0}
\titleformat{\section}{\large\bfseries}{\thesection}{1em}{}

\section{Estimated intervention effects}
{\renewcommand{\arraystretch}{0.6}
\begin{table}[!hb]
\caption{Bias of the estimated intervention effects from 1,000 simulations per parameter combination ($I= 8, 16, 32$; $K = 10, 100$; $J = 4, 8$) using four analysis models (Table~\ref{tab:mixed_models}; eq. = equation; cov. adj. = covariate adjustment), including only converged models. Closed and open cohort data with a covariate affecting the outcome linearly and nonlinearly.}
\label{tab:Bias}
\centering
\begin{tabular}{rp{5.8cm}rrrR{0.79cm}R{0.79cm}R{0.79cm}R{0.79cm}}
    \toprule
    \textbf{Eq.} & \textbf{Description of adjustment} & $\mathbf{I}$ & $\mathbf{J}$ & $\mathbf{K}$ &
    \multicolumn{2}{c}{\textbf{Closed cohort}} &
    \multicolumn{2}{c}{\textbf{Open cohort}} \\
    \cmidrule(lr){6-7} \cmidrule(lr){8-9}
    & & & & & \textbf{linear} &
    \parbox{0.8cm}{\centering\textbf{non-\\linear}} &
    \textbf{linear} &
    \parbox{0.8cm}{\centering\textbf{non-\\linear}} \\
    \midrule
   1 & Time-updated linear cov. adj. & 8 & 4 & 10 & 0.033 & -0.683 & 0.021 & 1.848\\
    1 & Time-updated linear cov. adj. & 8 & 4 & 100 & 0.004 & -0.743 & 0.010 & 2.289\\
    1 & Time-updated linear cov. adj. & 8 & 8 & 10 & -0.026 & -1.108 & -0.002 & 3.615\\
    1 & Time-updated linear cov. adj. & 8 & 8 & 100 & -0.008 & -1.023 & -0.009 & 3.958\\
    1 & Time-updated linear cov. adj. & 16 & 4 & 10 & -0.002 & -0.765 & -0.002 & 1.859\\
    1 & Time-updated linear cov. adj. & 16 & 4 & 100 & -0.002 & -0.736 & -0.001 & 2.297\\
    1 & Time-updated linear cov. adj. & 16 & 8 & 10 & 0.003 & -1.095 & 0.005 & 3.596\\
    1 & Time-updated linear cov. adj. & 16 & 8 & 100 & 0.009 & -1.046 & 0.003 & 3.962\\
    1 & Time-updated linear cov. adj. & 32 & 4 & 10 & -0.005 & -0.720 & -0.010 & 1.845\\
    1 & Time-updated linear cov. adj. & 32 & 4 & 100 & -0.000 & -0.737 & -0.000 & 2.294\\
    1 & Time-updated linear cov. adj. & 32 & 8 & 10 & 0.001 & -1.035 & 0.011 & 3.579\\
    1 & Time-updated linear cov. adj. & 32 & 8 & 100 & -0.002 & -1.042 & -0.002 & 3.963\\
    2 & Fixed time effects & 8 & 4 & 10 & 0.109 & -0.008 & 0.029 & 0.032\\
    2 & Fixed time effects & 8 & 4 & 100 & 0.004 & -0.014 & 0.007 & -0.012\\
    2 & Fixed time effects & 8 & 8 & 10 & -0.020 & -0.035 & -0.025 & 0.040\\
    2 & Fixed time effects & 8 & 8 & 100 & -0.010 & 0.020 & -0.002 & 0.003\\
    2 & Fixed time effects & 16 & 4 & 10 & 0.023 & -0.020 & -0.005 & 0.006\\
    2 & Fixed time effects & 16 & 4 & 100 & -0.006 & 0.004 & -0.002 & -0.002\\
    2 & Fixed time effects & 16 & 8 & 10 & -0.000 & -0.036 & 0.030 & 0.008\\
    2 & Fixed time effects & 16 & 8 & 100 & 0.005 & 0.012 & 0.005 & -0.003\\
    2 & Fixed time effects & 32 & 4 & 10 & 0.005 & -0.009 & -0.018 & -0.001\\
    2 & Fixed time effects & 32 & 4 & 100 & 0.004 & -0.001 & 0.001 & 0.005\\
    2 & Fixed time effects & 32 & 8 & 10 & 0.007 & -0.002 & 0.007 & 0.004\\
    2 & Fixed time effects & 32 & 8 & 100 & 0.003 & -0.003 & -0.001 & 0.002\\
    3 & Fixed time effects \& baseline cov. adj. & 8 & 4 & 10 & 0.074 & 0.035 & 0.018 & 0.024\\
    3 & Fixed time effects \& baseline cov. adj. & 8 & 4 & 100 & 0.002 & -0.013 & 0.008 & -0.008\\
    3 & Fixed time effects \& baseline cov. adj. & 8 & 8 & 10 & -0.021 & -0.126 & -0.025 & 0.067\\
    3 & Fixed time effects \& baseline cov. adj. & 8 & 8 & 100 & -0.015 & 0.023 & -0.008 & 0.005\\
    3 & Fixed time effects \& baseline cov. adj. & 16 & 4 & 10 & 0.004 & -0.015 & 0.003 & 0.008\\
    3 & Fixed time effects \& baseline cov. adj. & 16 & 4 & 100 & -0.006 & 0.009 & 0.003 & 0.001\\
    3 & Fixed time effects \& baseline cov. adj. & 16 & 8 & 10 & -0.006 & -0.017 & -0.002 & 0.003\\
    3 & Fixed time effects \& baseline cov. adj. & 16 & 8 & 100 & 0.005 & 0.004 & 0.007 & -0.005\\
    3 & Fixed time effects \& baseline cov. adj. & 32 & 4 & 10 & -0.011 &  0.010 & -0.013 & 0.004\\
    3 & Fixed time effects \& baseline cov. adj. & 32 & 4 & 100 & 0.004 & 0.001 & 0.003 & 0.003\\
    3 & Fixed time effects \& baseline cov. adj. & 32 & 8 & 10 & -0.003 & 0.027 & 0.009 & -0.011\\
    3 & Fixed time effects \& baseline cov. adj. & 32 & 8 & 100 & -0.000 & -0.003 & 0.003 & 0.001\\
    4 & Fixed time effects \& time-updated cov. adj. & 8 &  4 & 10 & 0.074 & 0.030 & 0.021 & 0.012\\
    4 & Fixed time effects \& time-updated cov. adj. & 8 &  4 & 100 & 0.003 & -0.014 & 0.009 & -0.009\\
    4 & Fixed time effects \& time-updated cov. adj. & 8 &  8 & 10 & -0.021 & -0.122 & -0.030 & 0.055\\
    4 & Fixed time effects \& time-updated cov. adj. & 8 &  8 & 100 & -0.009 & 0.018 & -0.010 & 0.000\\
    4 & Fixed time effects \& time-updated cov. adj. & 16 &  4 & 10 & 0.004 & -0.010 & -0.000 & 0.001\\
    4 & Fixed time effects \& time-updated cov. adj. & 16 &  4 & 100 & -0.007 & 0.010 & -0.001 & 0.003\\
    4 & Fixed time effects \& time-updated cov. adj. & 16 &  8 & 10 & -0.006 & -0.019 & 0.002 & 0.000\\
    4 & Fixed time effects \& time-updated cov. adj. & 16 &  8 & 100 & 0.011 & 0.002 & 0.005 & -0.001\\
    4 & Fixed time effects \& time-updated cov. adj. & 32 &  4 & 10 & -0.011 & 0.010 & -0.013 & 0.005\\
    4 & Fixed time effects \& time-updated cov. adj. & 32 &  4 & 100 & 0.004 & 0.001 & 0.004 & 0.004\\
    4 & Fixed time effects \& time-updated cov. adj. & 32 &  8 & 10 & -0.003 & 0.025 & 0.013 & -0.013\\
    4 & Fixed time effects \& time-updated cov. adj. & 32 &  8 & 100 & 0.007 & -0.005 & -0.000 & 0.002\\
    \bottomrule
\end{tabular}
\end{table}
\FloatBarrier

\begin{figure}[!htb]
    \centering
    \begin{subfigure}[b]{\textwidth}
        \centering\includegraphics[width=\textwidth]{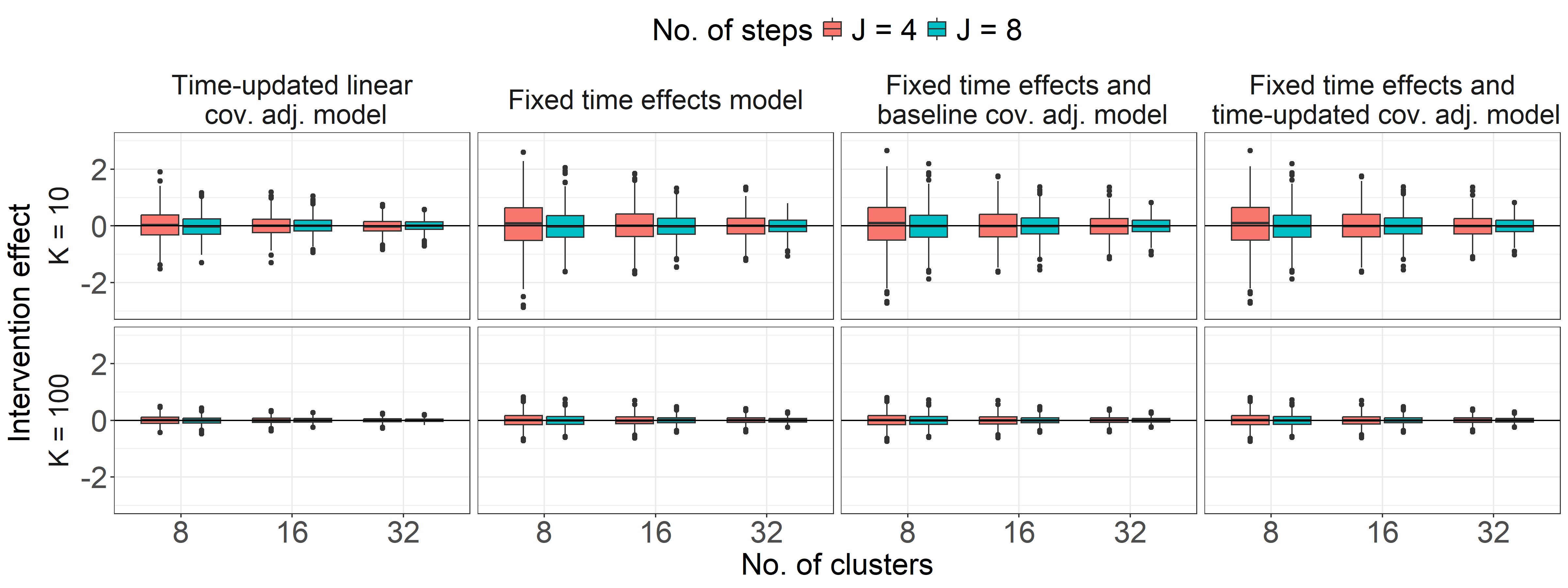}
        \caption{Linear influence}
        \label{subfig:Estimate_closedCohort_linear}
    \end{subfigure}
    \hfill
    \begin{subfigure}[b]{\textwidth}
        \centering\includegraphics[width=\textwidth]{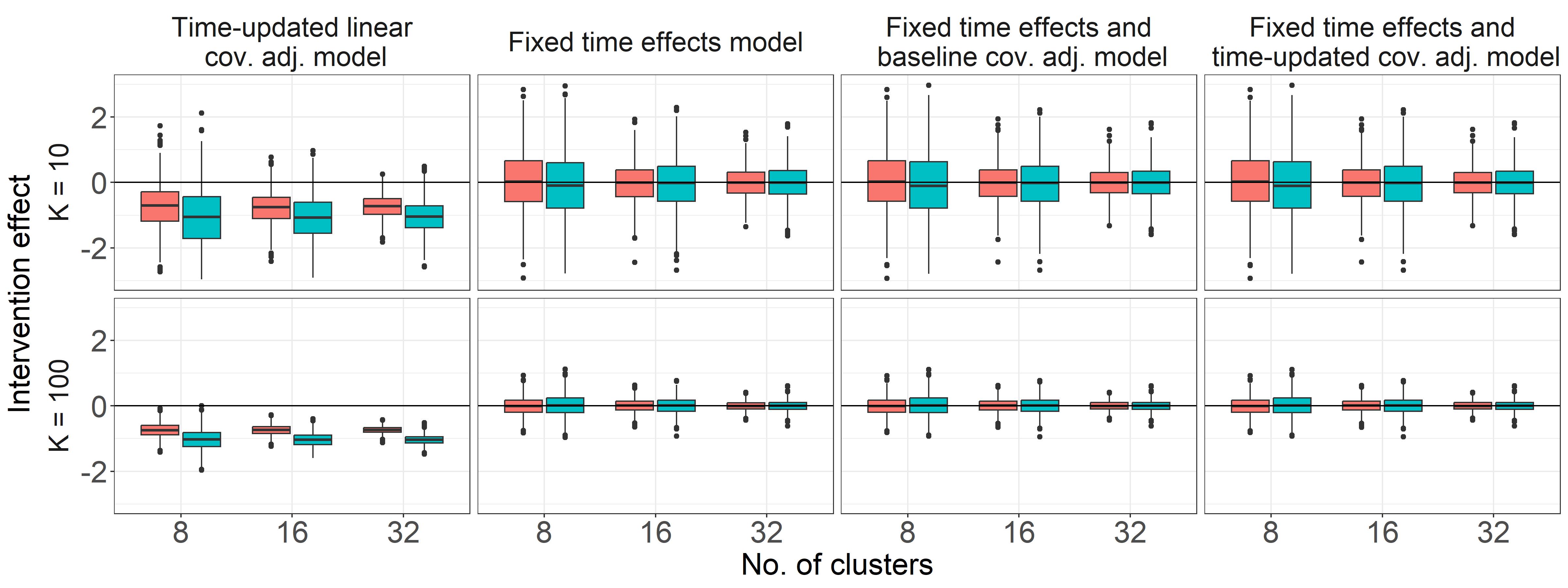}
        \caption{Nonlinear influence} \label{subfig:Estimate_closedCohort_nonlinear}
    \end{subfigure}
    \caption{Estimated intervention effects from 1,000 simulations per parameter combination ($I= 8, 16, 32$; $K = 10, 100$; $J = 4, 8$) using four analysis models (Table~\ref{tab:mixed_models}; cov. adj. = covariate adjustment). Closed cohort data with a covariate affecting the outcome (a) linearly and (b) nonlinearly. Results are shown from converged and non-converged models.}
    \label{fig:Estimate_closedCohort_withnonconverged}
\end{figure}
\FloatBarrier

\begin{figure}[!htb]
    \centering
    \begin{subfigure}[b]{\textwidth}
        \centering\includegraphics[width=0.9\textwidth]{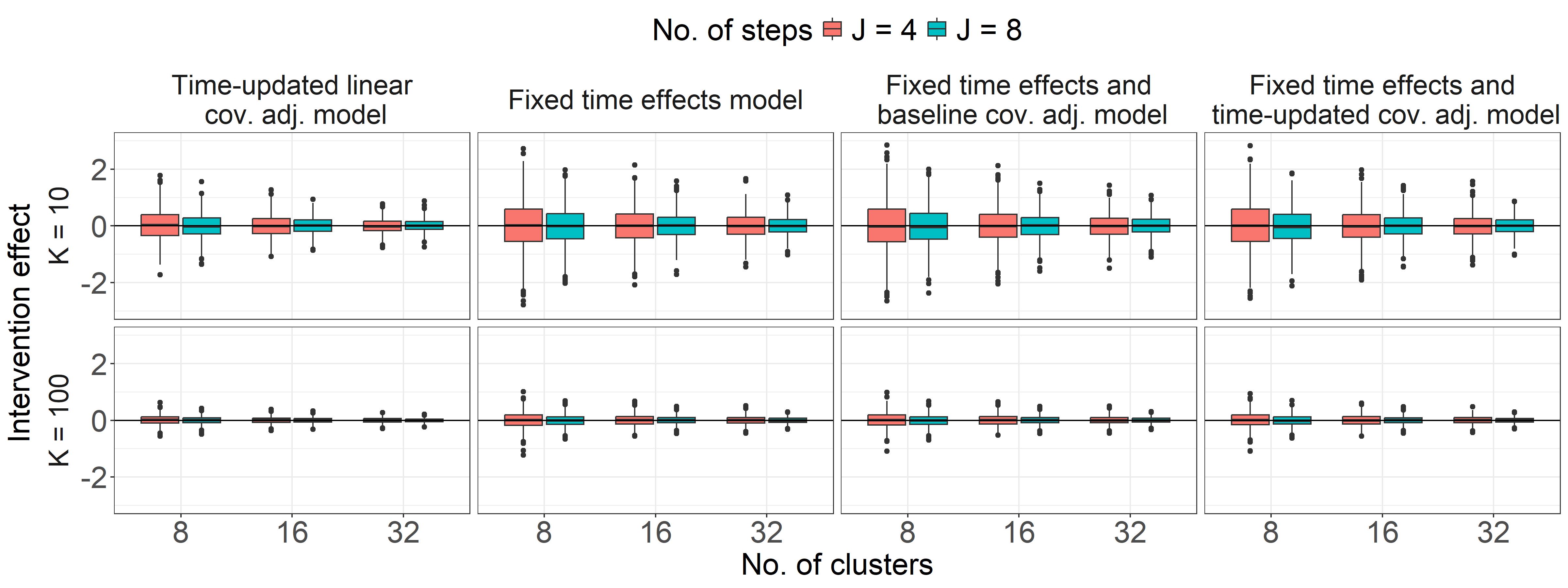}
        \caption{Linear influence}
        \label{fig:Estimate_openCohort_linear}
    \end{subfigure}
    \hfill
    \begin{subfigure}[b]{\textwidth}
        \centering\includegraphics[width=\textwidth]{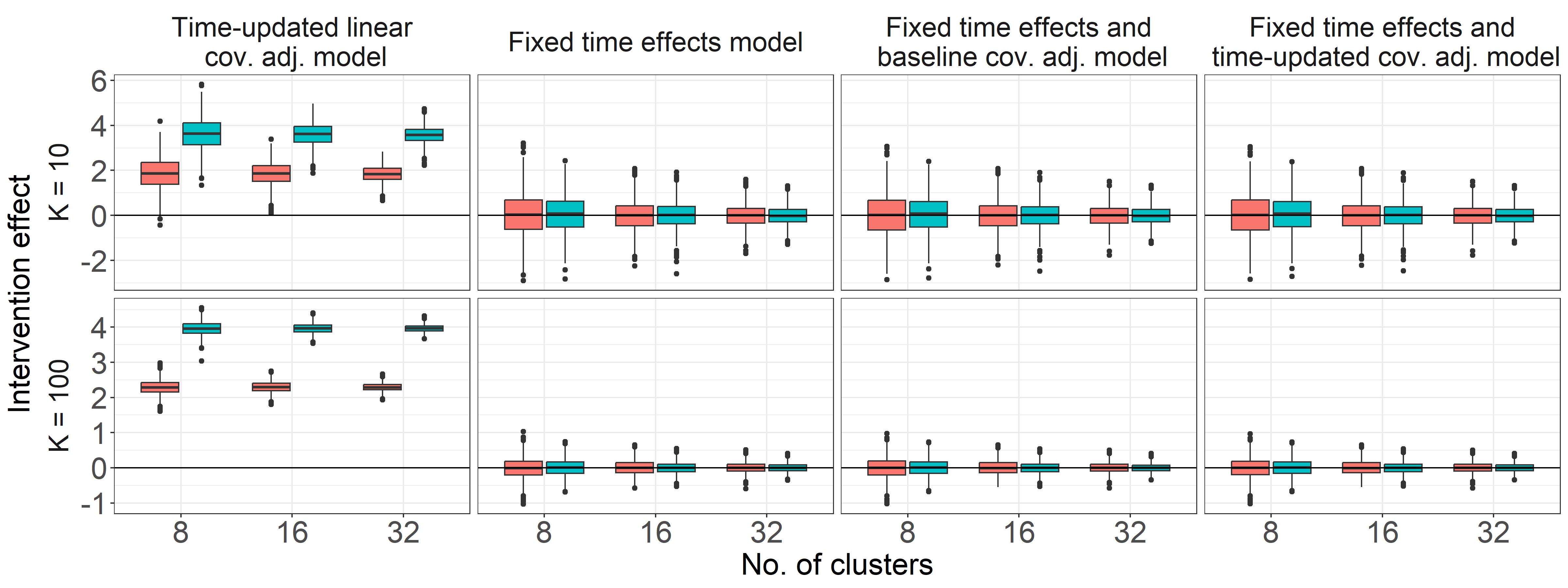}
        \caption{Nonlinear influence}
        \label{fig:Estimate_openCohort_nonlinear}
    \end{subfigure}
    \caption{Estimated intervention effects from 1,000 simulations per parameter combination ($I= 8, 16,$ and $32$, $K = 10,$ and $100, $J$ = 4,$ and $8$) using four analysis models (Table~\ref{tab:mixed_models}; cov. adj. = covariate adjustment). Open cohort data with a covariate affecting the outcome (a) linearly and (b) nonlinearly.}
    \label{fig:Estimate_openCohort}
\end{figure}
\FloatBarrier

\clearpage
\section{Coverage}

\subsection{Linear covariate effect}
\begin{table}[!htb]
\caption{Coverage derived from four analysis models (Table~\ref{tab:mixed_models}) without (standard) and with cluster-robust variance estimation, specifically CR0VE, CR2VE and CR3VE. Closed cohort data with $I = 8$ clusters, $J = 4$ steps, a cluster-period size of $K = 10$ or $K = 100$ and a linear covariate effect.}
\label{tab:Coverage_linear}
\centering
\begin{tabular}{p{0.2cm}p{0.2cm}p{0.2cm}p{1.8cm}R{1.2cm}R{1.2cm}R{1.2cm}R{1.2cm}}
    \toprule
    $\mathbf{I}$ & $\mathbf{J}$ & $\mathbf{K}$   & \textbf{Variance estimator} &
    \multicolumn{4}{c}{\textbf{Model}} \\
    \cmidrule(lr){5-8}
    & & & &
    \textbf{Eq. 1} &
    \textbf{Eq. 2} &
    \textbf{Eq. 3} &
    \textbf{Eq. 4} \\
    \midrule
     8 & 4 & 10 & Standard & 0.946 & 0.944 & 0.942 & 0.942\\
     8 & 4 & 10 & CR0 & 0.937 & 0.929 & 0.924 & 0.924\\
     8 & 4 & 10 & CR2 & 0.953 & 0.946 & 0.951 & 0.951\\
     8 & 4 & 10 & CR3 & 0.969 & 0.974 & 0.969 & 0.969\\
     16 & 4 & 10& Standard & 0.953 & 0.942 & 0.940 & 0.940\\
     16 & 4 & 10& CR0 & 0.953 & 0.939 & 0.926 & 0.926\\
     16 & 4 & 10& CR2 & 0.957 & 0.952 & 0.949 & 0.949\\
     16 & 4 & 10& CR3 & 0.963 & 0.959 & 0.956 & 0.956\\
     32 & 4 & 10& Standard & 0.938 & 0.954 & 0.954 & 0.954\\
     32 & 4 & 10& CR0 & 0.939 & 0.951 & 0.944 & 0.944\\
     32 & 4 & 10& CR2 & 0.941 & 0.957 & 0.948 & 0.948\\
     32 & 4 & 10& CR3 & 0.947 & 0.962 & 0.958 & 0.958\\
     8 & 4 & 100 & Standard & 0.954 & 0.950 & 0.952 & 0.952\\
     8 & 4 & 100 & CR0 & 0.941 & 0.926 & 0.927 & 0.927\\
     8 & 4 & 100 & CR2 & 0.958 & 0.959 & 0.955 & 0.955\\
     8 & 4 & 100 & CR3 & 0.969 & 0.977 & 0.979 & 0.979\\
     16 & 4 & 100& Standard & 0.957 & 0.946 & 0.948 & 0.948\\
     16 & 4 & 100& CR0 & 0.949 & 0.923 & 0.923 & 0.923\\
     16 & 4 & 100& CR2 & 0.955 & 0.943 & 0.943 & 0.943\\
     16 & 4 & 100& CR3 & 0.963 & 0.956 & 0.954 & 0.954\\
     32 & 4 & 100& Standard & 0.948 & 0.950 & 0.949 & 0.949\\
     32 & 4 & 100& CR0 & 0.935 & 0.945 & 0.946 & 0.946\\
     32 & 4 & 100& CR2 & 0.938 & 0.950 & 0.951 & 0.951\\
     32 & 4 & 100& CR3 & 0.945 & 0.955 & 0.956 & 0.956\\
     8 & 8 & 10 & Standard & 0.946 & 0.940 & 0.943 & 0.943\\
     8 & 8 & 10 & CR0 & 0.933 & 0.901 & 0.906 & 0.906\\
     8 & 8 & 10 & CR2 & 0.950 & 0.932 & 0.934 & 0.934\\
     8 & 8 & 10 & CR3 & 0.963 & 0.963 & 0.965 & 0.965\\
     16 & 8 & 10& Standard & 0.943 & 0.950 & 0.951 & 0.951\\
     16 & 8 & 10& CR0 & 0.940 & 0.939 & 0.937 & 0.937\\
     16 & 8 & 10& CR2 & 0.950 & 0.955 & 0.953 & 0.953\\
     16 & 8 & 10& CR3 & 0.956 & 0.965 & 0.962 & 0.962\\
     32 & 8 & 10& Standard & 0.955 & 0.951 & 0.947 & 0.947\\
     32 & 8 & 10& CR0 & 0.949 & 0.944 & 0.939 & 0.939\\
     32 & 8 & 10& CR2 & 0.953 & 0.954 & 0.952 & 0.952\\
     32 & 8 & 10& CR3 & 0.956 & 0.961 & 0.962 & 0.962\\
     8 & 8 & 100 & Standard & 0.941 & 0.946 & 0.946 & 0.947\\
     8 & 8 & 100 & CR0 & 0.929 & 0.907 & 0.907 & 0.906\\
     8 & 8 & 100 & CR2 & 0.945 & 0.954 & 0.954 & 0.953\\
     8 & 8 & 100 & CR3 & 0.953 & 0.974 & 0.974 & 0.974\\
     16 & 8 & 100& Standard & 0.947 & 0.954 & 0.953 & 0.953\\
     16 & 8 & 100& CR0 & 0.940 & 0.938 & 0.939 & 0.939\\
     16 & 8 & 100& CR2 & 0.945 & 0.948 & 0.950 & 0.950\\
     16 & 8 & 100& CR3 & 0.953 & 0.957 & 0.957 & 0.957\\
     32 & 8 & 100& Standard & 0.939 & 0.931 & 0.930 & 0.930\\
     32 & 8 & 100& CR0 & 0.944 & 0.930 & 0.931 & 0.931\\
     32 & 8 & 100& CR2 & 0.947 & 0.941 & 0.942 & 0.941\\
     32 & 8 & 100& CR3 & 0.951 & 0.948 & 0.951 & 0.950\\
    \bottomrule
\end{tabular}
\end{table}
\FloatBarrier
\clearpage

\subsection{Nonlinear covariate effect}

\begin{table}[!htb]
\caption{Coverage derived from four analysis models (Table~\ref{tab:mixed_models}) without (standard) and with cluster-robust variance estimation, specifically CR0VE, CR2VE and CR3VE. Closed cohort data with $I = 8$ clusters, $J = 4$ steps, a cluster-period size of $K = 10$ or $K = 100$ and a nonlinear covariate effect.}
\label{tab:PercSEBias_nonlinear}
\centering
\begin{tabular}{p{0.2cm}R{0.2cm}R{0.2cm}p{1.8cm}R{1.2cm}R{1.2cm}R{1.2cm}R{1.2cm}}
    \toprule
    $\mathbf{I}$ & $\mathbf{J}$ & $\mathbf{K}$   & \textbf{Variance estimator} &
    \multicolumn{4}{c}{\textbf{Model}} \\
    \cmidrule(lr){5-8}
    & & & &
    \textbf{Eq. 1} &
    \textbf{Eq. 2} &
    \textbf{Eq. 3} &
    \textbf{Eq. 4} \\
    \midrule
     8 & 4 & 10 & Standard & 0.841 & 0.977 & 0.975 & 0.975\\
     8 & 4 & 10 & CR0 & 0.830 & 0.920 & 0.922 & 0.922\\
     8 & 4 & 10 & CR2 & 0.851 & 0.944 & 0.943 & 0.943\\
     8 & 4 & 10 & CR3 & 0.878 & 0.969 & 0.972 & 0.972\\
     16 & 4 & 10& Standard & 0.677 & 0.989 & 0.990 & 0.990\\
     16 & 4 & 10& CR0 & 0.671 & 0.950 & 0.952 & 0.952\\
     16 & 4 & 10& CR2 & 0.693 & 0.962 & 0.960 & 0.960\\
     16 & 4 & 10& CR3 & 0.710 & 0.970 & 0.971 & 0.971\\
     32 & 4 & 10& Standard & 0.473 & 0.977 & 0.974 & 0.974\\
     32 & 4 & 10& CR0 & 0.450 & 0.939 & 0.940 & 0.940\\
     32 & 4 & 10& CR2 & 0.463 & 0.948 & 0.945 & 0.945\\
     32 & 4 & 10& CR3 & 0.474 & 0.954 & 0.951 & 0.951\\
     8 & 4 & 100 & Standard & 0.078 & 0.985 & 0.985 & 0.985\\
     8 & 4 & 100 & CR0 & 0.124 & 0.932 & 0.932 & 0.932\\
     8 & 4 & 100 & CR2 & 0.170 & 0.959 & 0.959 & 0.959\\
     8 & 4 & 100 & CR3 & 0.204 & 0.976 & 0.976 & 0.976\\
     16 & 4 & 100& Standard & 0.003 & 0.978 & 0.978 & 0.978\\
     16 & 4 & 100& CR0 & 0.005 & 0.929 & 0.929 & 0.929\\
     16 & 4 & 100& CR2 & 0.006 & 0.949 & 0.949 & 0.949\\
     16 & 4 & 100& CR3 & 0.009 & 0.957 & 0.957 & 0.957\\
     32 & 4 & 100& Standard & 0.000 & 0.979 & 0.978 & 0.978\\
     32 & 4 & 100& CR0 & 0.000 & 0.943 & 0.947 & 0.947\\
     32 & 4 & 100& CR2 & 0.000 & 0.945 & 0.950 & 0.950\\
     32 & 4 & 100& CR3 & 0.000 & 0.956 & 0.960 & 0.960\\
     8 & 8 & 10 & Standard & 0.731 & 0.963 & 0.962 & 0.962\\
     8 & 8 & 10 & CR0 & 0.800 & 0.930 & 0.931 & 0.931\\
     8 & 8 & 10 & CR2 & 0.837 & 0.958 & 0.962 & 0.962\\
     8 & 8 & 10 & CR3 & 0.867 & 0.985 & 0.980 & 0.980\\
     16 & 8 & 10& Standard & 0.574 & 0.943 & 0.939 & 0.939\\
     16 & 8 & 10& CR0 & 0.647 & 0.926 & 0.931 & 0.931\\
     16 & 8 & 10& CR2 & 0.675 & 0.941 & 0.945 & 0.945\\
     16 & 8 & 10& CR3 & 0.712 & 0.964 & 0.962 & 0.962\\
     32 & 8 & 10& Standard & 0.348 & 0.955 & 0.959 & 0.959\\
     32 & 8 & 10& CR0 & 0.428 & 0.944 & 0.942 & 0.942\\
     32 & 8 & 10& CR2 & 0.444 & 0.952 & 0.952 & 0.952\\
     32 & 8 & 10& CR3 & 0.465 & 0.962 & 0.959 & 0.959\\
     8 & 8 & 100 & Standard & 0.047 & 0.943 & 0.942 & 0.942\\
     8 & 8 & 100 & CR0 & 0.160 & 0.922 & 0.925 & 0.925\\
     8 & 8 & 100 & CR2 & 0.223 & 0.953 & 0.953 & 0.953\\
     8 & 8 & 100 & CR3 & 0.298 & 0.971 & 0.968 & 0.968\\
     16 & 8 & 100& Standard & 0.000 & 0.952 & 0.951 & 0.951\\
     16 & 8 & 100& CR0 & 0.004 & 0.936 & 0.933 & 0.933\\
     16 & 8 & 100& CR2 & 0.006 & 0.944 & 0.946 & 0.946\\
     16 & 8 & 100& CR3 & 0.009 & 0.968 & 0.964 & 0.964\\
     32 & 8 & 100& Standard & 0.000 & 0.958 & 0.958 & 0.958\\
     32 & 8 & 100& CR0 & 0.000 & 0.943 & 0.943 & 0.943\\
     32 & 8 & 100& CR2 & 0.000 & 0.948 & 0.948 & 0.948\\
     32 & 8 & 100& CR3 & 0.000 & 0.961 & 0.961 & 0.961\\
    \bottomrule
\end{tabular}
\end{table}
\FloatBarrier
\clearpage

\section{Percentage standard error}

\begin{figure}[!htb]
    \centering
    \begin{subfigure}[b]{0.49\textwidth}
        \includegraphics[width=\textwidth]{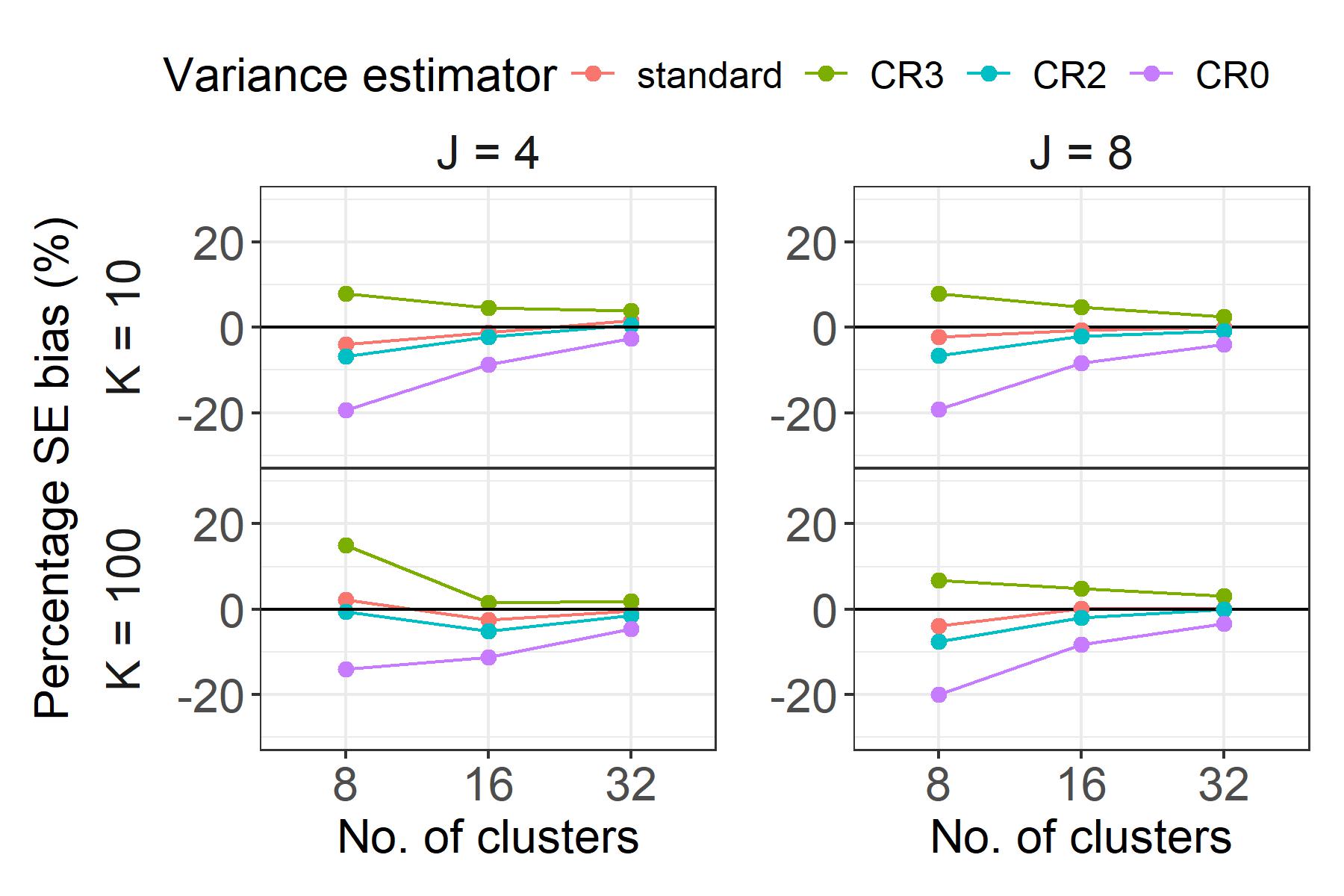}
        \caption{Closed cohort and linear influence}
        \label{fig:SEBias_closedCohort_linear}
    \end{subfigure}
    \hfill
    \begin{subfigure}[b]{0.49\textwidth}
        \includegraphics[width=\textwidth]{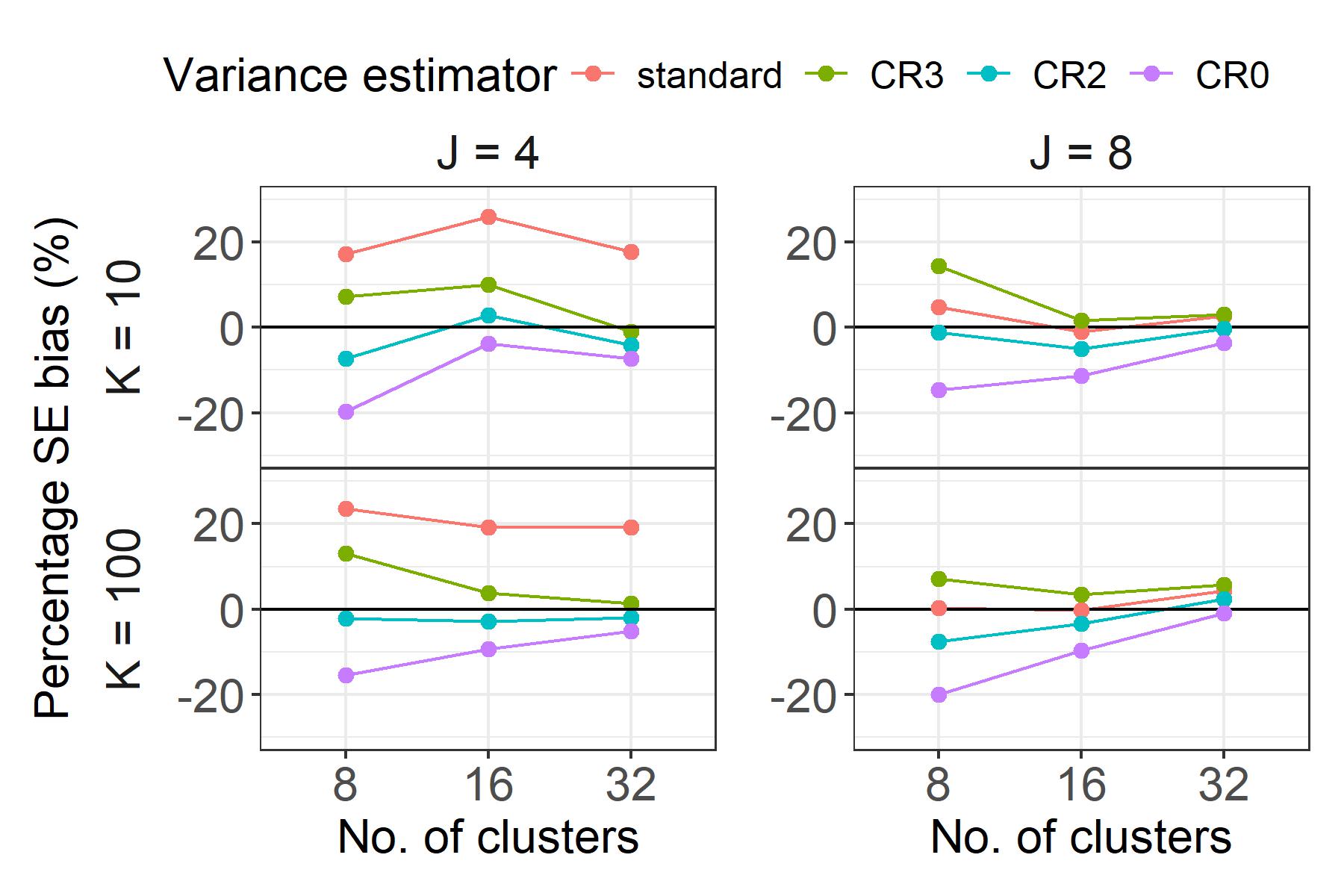}
        \caption{Closed cohort and nonlinear influence}
        \label{fig:SEBias_closedCohort_nonlinear}
    \end{subfigure}
    \centering
    \begin{subfigure}[b]{0.49\textwidth}
        \includegraphics[width=\textwidth]{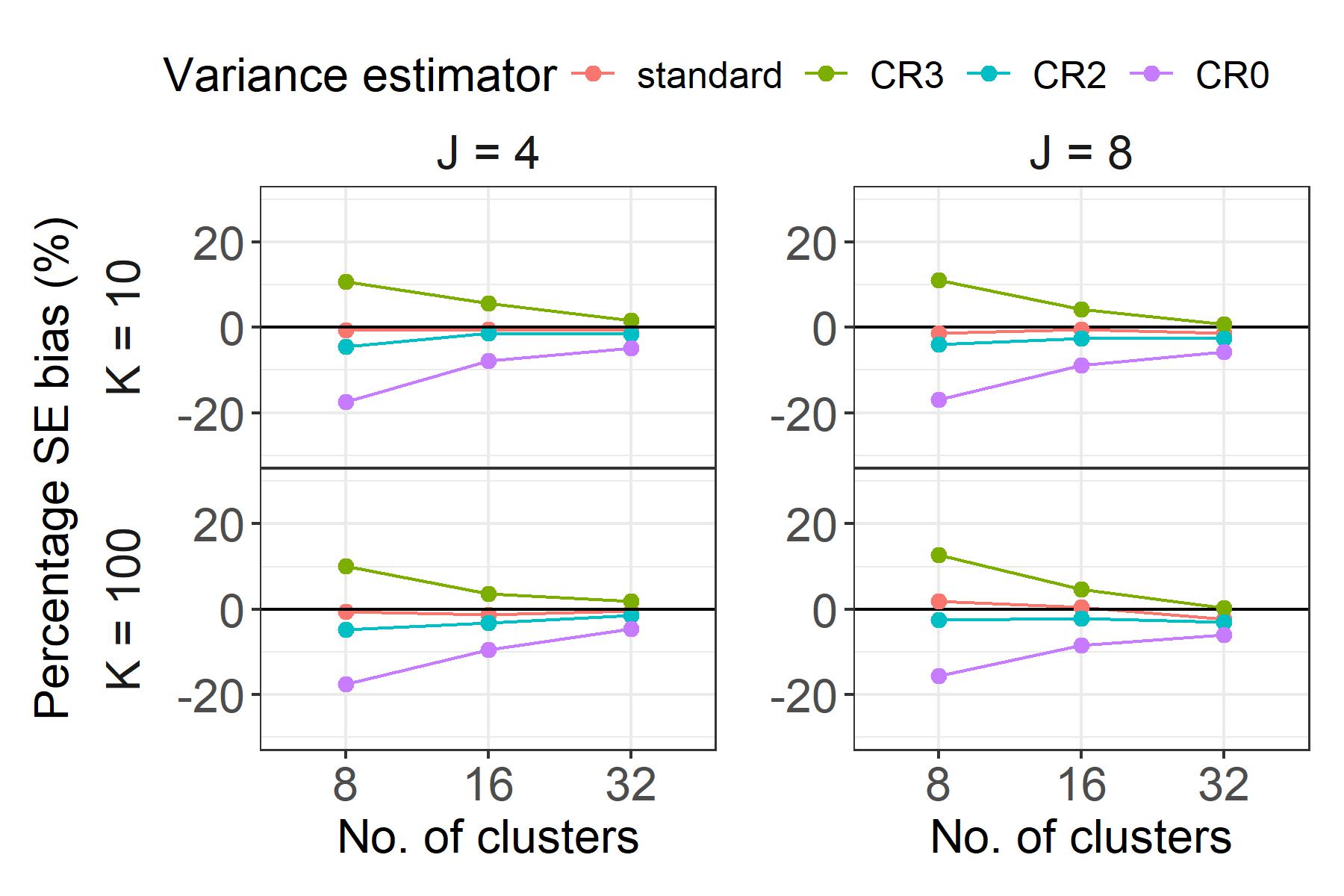}
        \caption{Open cohort and linear influence}
        \label{fig:SEBias_openCohort_linear}
    \end{subfigure}
    \hfill
    \begin{subfigure}[b]{0.49\textwidth}
        \includegraphics[width=\textwidth]{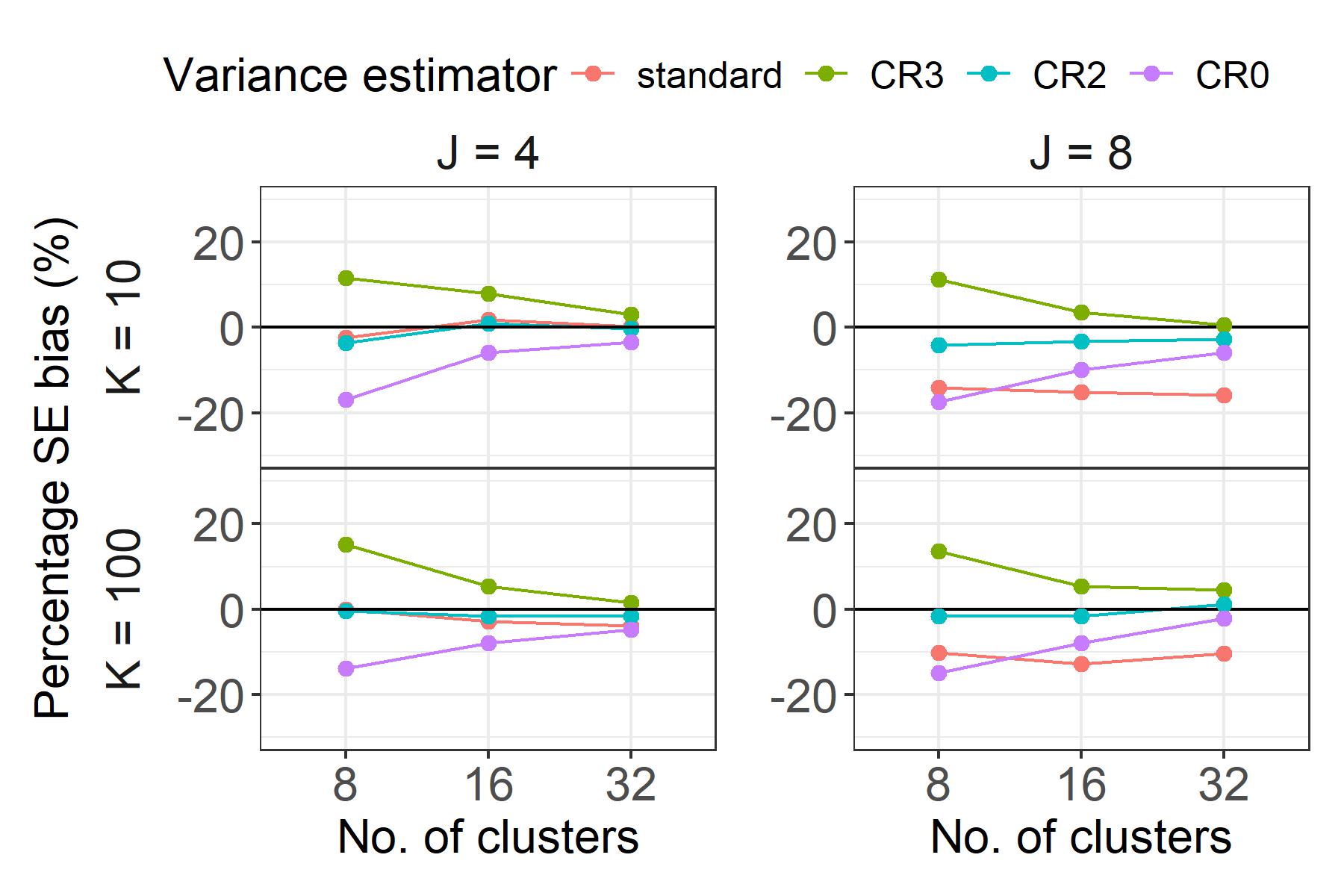}
        \caption{Open cohort and nonlinear influence}
        \label{fig:SEBias_openCohort_nonlinear}
    \end{subfigure}
    \caption{Percentage error of model-based standard errors (SE) compared to empirical SE from 1,000 simulations per parameter combination ($I = 8, 16, 32; K = 10, 100; J = 4, 8$). The analysis model includes fixed categorical time effects (Equation 2) and is presented with standard estimation (red line) and three cluster-robust variance estimation methods: CR0VE (purple line), CR2VE (blue line), and CR3VE (green line). Results are displayed for four scenarios: (a) closed cohort with linear covariate influence, (b) closed cohort with nonlinear covariate influence, (c) open cohort with linear covariate influence, and (d) open cohort with nonlinear covariate influence.}
    \label{fig:SEBias_all}
\end{figure}
\FloatBarrier

Figure~\ref{fig:SEBias_all} shows that the averaged model-based SEs derived from models employing standard variance estimation (without cluster-robust variance estimation; red line) accurately estimate the true SEs when the time-varying covariate linearly influences the outcome in both closed and open cohort settings (Subfigs.~\ref{fig:SEBias_closedCohort_linear} and~\ref{fig:SEBias_openCohort_linear}, respectively). However, these average model-based SEs may be biased if the time-varying covariate's influence on the outcome is nonlinear (Subfigs.~\ref{fig:SEBias_closedCohort_nonlinear} and~\ref{fig:SEBias_openCohort_nonlinear}). Depending on the parameter combination, this bias results in either an overestimation of up to 26\% (observed for $J = 4$ in closed cohort settings) or an underestimation of up to 16\% (observed for $J = 8$ in open cohort settings) in the average model-based SE compared to the empirical SE of the estimated intervention effects derived from the simulation.
\par
Across all parameter combinations and settings in our simulation, the average model-based SEs from the model with CR0 variance estimation (purple line) are consistently underestimated compared to the empirical SEs. Conversely, the model-based SEs from the models using the CR3VE (green line) slightly overestimate the true SEs, especially when the number of clusters is small. The model-based SEs from models with CR2 variance estimation (blue line) had the smallest relative errors compared to other types of CRVEs, although the error could still be as high as 8\%.

\subsection{Linear covariate effect}
\begin{table}[!htb]
\caption{Percentage standard error bias derived from four analysis models (Table~\ref{tab:mixed_models}) without (standard) and with cluster-robust variance estimation, specifically CR0VE, CR2VE and CR3VE. Closed cohort data with $I = 8$ clusters, $J = 4$ steps, a cluster-period size of $K = 10$ or $K = 100$ and a linear covariate effect.}
\label{tab:PercSEBias_linear}
\centering
\begin{tabular}{p{0.2cm}p{0.2cm}p{0.2cm}p{1.8cm}R{1.2cm}R{1.2cm}R{1.2cm}R{1.2cm}}
    \toprule
    $\mathbf{I}$ & $\mathbf{J}$ & $\mathbf{K}$   & \textbf{Variance estimator} &
    \multicolumn{4}{c}{\textbf{Model}} \\
    \cmidrule(lr){5-8}
    & & & &
    \textbf{Eq. 1} &
    \textbf{Eq. 2} &
    \textbf{Eq. 3} &
    \textbf{Eq. 4} \\
    \midrule
     8 & 4 & 10 & Standard & -1.457 & -3.976 & -4.327 & -4.327\\
     8 & 4 & 10 & CR0 & -10.594 & -19.319 & -19.764 & -19.764\\
     8 & 4 & 10 & CR2 & -4.137 & -6.735 & -7.208 & -7.208\\
     8 & 4 & 10 & CR3 & 2.847 & 7.844 & 7.368 & 7.368\\
     16 & 4 & 10& Standard & 2.577 & -1.119 & -1.837 & -1.837\\
     16 & 4 & 10& CR0 & -2.456 & -8.629 & -9.023 & -9.023\\
     16 & 4 & 10& CR2 & 0.892 & -2.263 & -2.654 & -2.654\\
     16 & 4 & 10& CR3 & 4.361 & 4.552 & 4.145 & 4.145\\
     32 & 4 & 10& Standard & -2.203 & 1.668 & 1.442 & 1.442\\
     32 & 4 & 10& CR0 & -5.324 & -2.597 & -2.919 & -2.919\\
     32 & 4 & 10& CR2 & -3.736 & 0.628 & 0.301 & 0.301\\
     32 & 4 & 10& CR3 & -2.120 & 3.961 & 3.632 & 3.632\\
     8 & 4 & 100 & Standard & -0.704 & 2.224 & 1.839 & 1.872\\
     8 & 4 & 100 & CR0 & -10.427 & -13.973 & -14.302 & -14.276\\
     8 & 4 & 100 & CR2 & -3.985 & -0.518 & -0.895 & -0.865\\
     8 & 4 & 100 & CR3 & 2.960 & 15.065 & 14.635 & 14.670\\
     16 & 4 & 100& Standard & 0.284 & -2.485 & -2.535 & -2.531\\
     16 & 4 & 100& CR0 & -4.801 & -11.246 & -11.253 & -11.251\\
     16 & 4 & 100& CR2 & -1.556 & -5.060 & -5.064 & -5.062\\
     16 & 4 & 100& CR3 & 1.808 & 1.563 & 1.558 & 1.560\\
     32 & 4 & 100& Standard & -2.259 & -0.382 & -0.307 & -0.306\\
     32 & 4 & 100& CR0 & -4.855 & -4.553 & -4.424 & -4.424\\
     32 & 4 & 100& CR2 & -3.271 & -1.395 & -1.260 & -1.259\\
     32 & 4 & 100& CR3 & -1.660 & 1.870 & 2.009 & 2.009\\
     8 & 8 & 10 & Standard & -0.192 & -2.165 & -2.236 & -2.236\\
     8 & 8 & 10 & CR0 & -11.372 & -19.113 & -19.142 & -19.142\\
     8 & 8 & 10 & CR2 & -4.642 & -6.544 & -6.547 & -6.547\\
     8 & 8 & 10 & CR3 & 2.674 & 7.989 & 8.011 & 8.011\\
     16 & 8 & 10& Standard & -2.247 & -0.604 & -1.354 & -1.354\\
     16 & 8 & 10& CR0 & -6.306 & -8.358 & -8.990 & -8.990\\
     16 & 8 & 10& CR2 & -2.937 & -2.009 & -2.671 & -2.671\\
     16 & 8 & 10& CR3 & 0.574 & 4.784 & 4.084 & 4.084\\
     32 & 8 & 10& Standard & -0.588 & -0.040 & -0.079 & -0.079\\
     32 & 8 & 10& CR0 & -2.688 & -3.961 & -4.067 & -4.067\\
     32 & 8 & 10& CR2 & -0.981 & -0.799 & -0.906 & -0.906\\
     32 & 8 & 10& CR3 & 0.760 & 2.467 & 2.362 & 2.362\\
     8 & 8 & 100 & Standard & -3.301 & -3.903 & -4.006 & -3.995\\
     8 & 8 & 100 & CR0 & -14.434 & -19.960 & -20.019 & -20.010\\
     8 & 8 & 100 & CR2 & -7.972 & -7.525 & -7.589 & -7.579\\
     8 & 8 & 100 & CR3 & -0.954 & 6.854 & 6.780 & 6.792\\
     16 & 8 & 100& Standard & -0.831 & 0.093 & 0.118 & 0.103\\
     16 & 8 & 100& CR0 & -5.738 & -8.272 & -8.307 & -8.319\\
     16 & 8 & 100& CR2 & -2.355 & -1.911 & -1.951 & -1.964\\
     16 & 8 & 100& CR3 & 1.165 & 4.891 & 4.851 & 4.838\\
     32 & 8 & 100& Standard & -2.648 & 0.118 & 0.220 & 0.225\\
     32 & 8 & 100& CR0 & -4.484 & -3.293 & -3.200 & -3.193\\
     32 & 8 & 100& CR2 & -2.819 & -0.110 & -0.013 & -0.006\\
     32 & 8 & 100& CR3 & -1.121 & 3.179 & 3.279 & 3.287\\
    \bottomrule
\end{tabular}
\end{table}
\FloatBarrier
\clearpage

\subsection{Nonlinear covariate effect}

\begin{table}[!htb]
\caption{Percentage standard error bias derived from four analysis models (Table~\ref{tab:mixed_models}) without (standard) and with cluster-robust variance estimation, specifically CR0VE, CR2VE and CR3VE. Closed cohort data with $I = 8$ clusters, $J = 4$ steps, a cluster-period size of $K = 10$ or $K = 100$ and a nonlinear covariate effect.}
\label{tab:PercSEBias_nonlinear}
\centering
\begin{tabular}{p{0.2cm}R{0.2cm}R{0.2cm}p{1.8cm}R{1.2cm}R{1.2cm}R{1.2cm}R{1.2cm}}
    \toprule
    $\mathbf{I}$ & $\mathbf{J}$ & $\mathbf{K}$   & \textbf{Variance estimator} &
    \multicolumn{4}{c}{\textbf{Model}} \\
    \cmidrule(lr){5-8}
    & & & &
    \textbf{Eq. 1} &
    \textbf{Eq. 2} &
    \textbf{Eq. 3} &
    \textbf{Eq. 4} \\
    \midrule
     8 & 4 & 10 & Standard & 4.068 & 17.199 & 17.002 & 17.002\\
     8 & 4 & 10 & CR0 & -12.474 & -19.780 & -19.902 & -19.691\\
     8 & 4 & 10 & CR2 & -3.100 & -7.262 &- 7.111 & -7.111\\
     8 & 4 & 10 & CR3 & 3.953 & 7.243 & 7.404 & 7.404\\
     16 & 4 & 10& Standard & 3.414 & 25.832 & 25.761 & 25.761\\
     16 & 4 & 10& CR0 & -4.936 & -5.282 & -5.028 & -5.245\\
     16 & 4 & 10& CR2 & -1.560 & 2.832 & 2.871 & 2.871\\
     16 & 4 & 10& CR3 & 1.843 & 9.987 & 10.030 & 10.030\\
     32 & 4 & 10& Standard & 6.179 & 17.705 & 16.975 & 16.975\\
     32 & 4 & 10& CR0 & 1.094 & -9.797 & -5.970 & -6.016\\
     32 & 4 & 10& CR2 & 1.179 & -4.215 & -4.680 & -4.680\\
     32 & 4 & 10& CR3 & 2.871 & -1.050 & -1.531 & -1.531\\
     8 & 4 & 100 & Standard & 5.508 & 23.553 & 23.449 & 23.417\\
     8 & 4 & 100 & CR0 & -9.725 & -15.529 & -14.891 & -14.623\\
     8 & 4 & 100 & CR2 & -2.667 & -2.180 & -2.029 & -2.054\\
     8 & 4 & 100 & CR3 & 4.503 & 13.103 & 13.260 & 13.231\\
     16 & 4 & 100& Standard & 3.903 & 19.112 & 18.794 & 18.810\\
     16 & 4 & 100& CR0 & -3.792 & -12.330 & -10.119 & -9.578\\
     16 & 4 & 100& CR2 & -0.650 & -2.894 & -3.104 & -3.096\\
     16 & 4 & 100& CR3 & 2.793 & 3.869 & 3.643 & 3.652\\
     32 & 4 & 100& Standard & 4.644 & 19.267 & 19.464 & 19.481\\
     32 & 4 & 100& CR0 & -0.687 & -4.024 & -4.944 & -5.404\\
     32 & 4 & 100& CR2 & 1.091 & -1.933 & -1.664 & -1.650\\
     32 & 4 & 100& CR3 & 2.814 & 1.307 & 1.582 & 1.597\\
     8 & 8 & 10 & Standard & -9.523 & 4.700 & 4.609 & 4.609\\
     8 & 8 & 10 & CR0 & -6.542 & -23.982 & -13.531 & -13.526\\
     8 & 8 & 10 & CR2 & -0.537 & -1.193 & -1.212 & -1.212\\
     8 & 8 & 10 & CR3 & 8.152 & 14.304 & 14.306 & 14.306\\
     16 & 8 & 10& Standard & -11.716 & -0.985 & -1.161 & -1.161\\
     16 & 8 & 10& CR0 & -6.842 & -18.900 & -12.678 & -12.816\\
     16 & 8 & 10& CR2 & -3.874 & -5.056 & -5.187 & -5.187\\
     16 & 8 & 10& CR3 & 0.055 & 1.616 & 1.487 & 1.487\\
     32 & 8 & 10& Standard & -10.552 & 2.727 & 2.696 & 2.696\\
     32 & 8 & 10& CR0 & -4.027 & -6.361 & -3.022 & -2.997\\
     32 & 8 & 10& CR2 & -0.585 & -0.348 & -0.276 & -0.276\\
     32 & 8 & 10& CR3 & 1.383 & 2.976 & 3.057 & 3.057\\
     8 & 8 & 100 & Standard & -9.718 & 0.332 & 0.106 & 0.107\\
     8 & 8 & 100 & CR0 & -0.268 & -24.059 & -17.478 & -18.131\\
     8 & 8 & 100 & CR2 & 7.365 & -7.482 & -7.611 & -7.609\\
     8 & 8 & 100 & CR3 & 16.745 & 7.060 & 6.915 & 6.917\\
     16 & 8 & 100& Standard & -7.406 & -0.201 & -0.606 & -0.608\\
     16 & 8 & 100& CR0 & 8.814 & -12.915 & -9.601 & -9.255\\
     16 & 8 & 100& CR2 & 11.127 & -3.290 & -3.556 & -3.557\\
     16 & 8 & 100& CR3 & 17.853 & 3.493 & 3.210 & 3.209\\
     32 & 8 & 100& Standard & -9.058 & 4.294 & 4.294 & 4.293\\
     32 & 8 & 100& CR0 & 8.665 & -1.414 & -1.414 & -1.319\\
     32 & 8 & 100& CR2 & 13.161 & 2.349 & 2.349 & 2.348\\
     32 & 8 & 100& CR3 & 13.369 & 5.765 & 5.765 & 5.765\\
    \bottomrule
\end{tabular}
\end{table}
\FloatBarrier
\clearpage

\section{Statistical power}

\subsection{Linear covariate effect}
\begin{figure}[!htb]
    \centerline{\includegraphics[width=\textwidth]{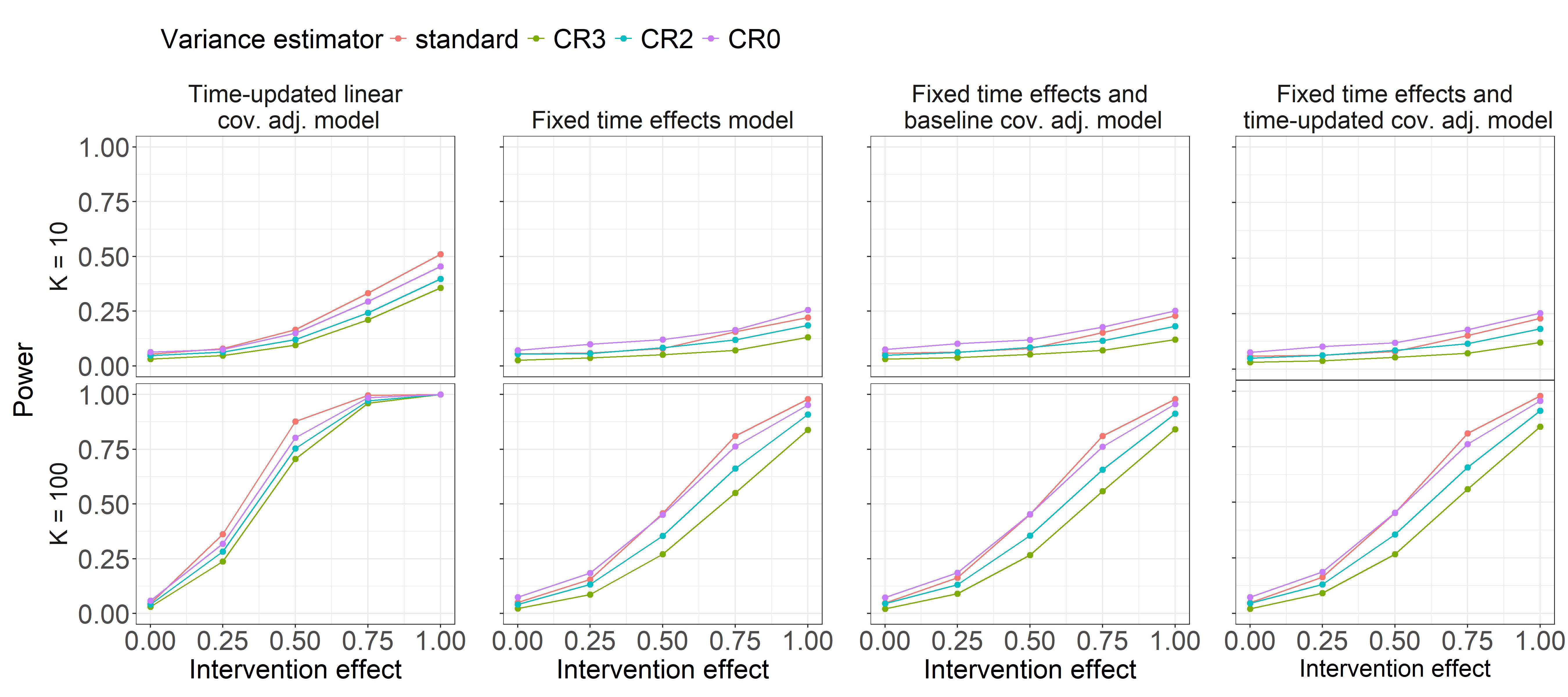}}
    \caption{Statistical power derived from four analysis model (Table~\ref{tab:mixed_models}) without (standard, red line) and with cluster-robust variance estimation, specifically CR0VE (purple line), CR2VE (blue line) and CR3VE (green line). Closed cohort data with $I = 8$ clusters, $J = 4$ steps, a cluster-period size of $K = 10$ (top) or $K = 100$ (bottom) and a linear covariate effect. The statistical power analysis was conducted via simulation, involving dataset generation, hypothesis testing, and calculation of the significance ratio.\label{fig:Power_closedCohort_linear}}
\end{figure}
\FloatBarrier

\begin{table}[!htb]
\caption{Statistical power derived from four analysis models (Table~\ref{tab:mixed_models}) without (standard) and with cluster-robust variance estimation, specifically CR0VE, CR2VE and CR3VE. Closed cohort data with $I = 8$ clusters, $J = 4$ steps, a cluster-period size of $K = 10$ or $K = 100$ and a linear covariate effect.}
\label{tab:Power_linear}
\centering
\begin{tabular}{R{0.2cm}p{0.3cm}p{1.8cm}R{1.2cm}R{1.2cm}R{1.2cm}R{1.2cm}}
    \toprule
    $\mathbf{K}$ & $\boldsymbol{\theta}$ & \textbf{Variance estimator} &
    \multicolumn{4}{c}{\textbf{Model}} \\
    \cmidrule(lr){4-7}
    & & &
    \textbf{Eq. 1} &
    \textbf{Eq. 2} &
    \textbf{Eq. 3} &
    \textbf{Eq. 4} \\
    \midrule
     10 & 0 & Standard & 0.054 & 0.056 & 0.058 & 0.058\\
     10 & 0 & CR0 & 0.063 & 0.071 & 0.076 & 0.076\\
     10 & 0 & CR2 & 0.047 & 0.054 & 0.049 & 0.049\\
     10 & 0 & CR3 & 0.031 & 0.026 & 0.031 & 0.031\\
     10 & 0.25 & Standard & 0.079 & 0.059 & 0.063 & 0.063\\
     10 & 0.25 & CR0 & 0.076 & 0.099 & 0.102 & 0.102\\
     10 & 0.25 & CR2 & 0.064 & 0.057 & 0.062 & 0.062\\
     10 & 0.25 & CR3 & 0.048 & 0.037 & 0.038 & 0.038\\
     10 & 0.5 & Standard & 0.166 & 0.079 & 0.080 & 0.080\\
     10 & 0.5 & CR0 & 0.149 & 0.120 & 0.119 & 0.119\\
     10 & 0.5 & CR2 & 0.120 & 0.083 & 0.085 & 0.085\\
     10 & 0.5 & CR3 & 0.095 & 0.052 & 0.053 & 0.053\\
     10 & 0.75 & Standard & 0.332 & 0.156 & 0.152 & 0.152\\
     10 & 0.75 & CR0 & 0.294 & 0.164 & 0.177 & 0.177\\
     10 & 0.75 & CR2 & 0.243 & 0.119 & 0.115 & 0.115\\
     10 & 0.75 & CR3 & 0.210 & 0.071 & 0.072& 0.072\\
     10 & 1 & Standard & 0.511 & 0.221 & 0.229 & 0.229\\
     10 & 1 & CR0 & 0.454 & 0.256 & 0.252 & 0.252\\
     10 & 1 & CR2 & 0.397 & 0.186 & 0.182 & 0.182\\
     10 & 1 & CR3 & 0.356 & 0.131 & 0.120 & 0.120\\
     100 & 0 & Standard & 0.046 & 0.050 & 0.048 & 0.048\\
     100 & 0 & CR0 & 0.059 & 0.074 & 0.073 & 0.073\\
     100 & 0 & CR2 & 0.042 & 0.041 & 0.045 & 0.045\\
     100 & 0 & CR3 & 0.031 & 0.023 & 0.021 & 0.021\\
     100 & 0.25 & Standard & 0.362 & 0.156 & 0.163 & 0.163\\
     100 & 0.25 & CR0 & 0.319 & 0.185 & 0.186 & 0.186\\
     100 & 0.25 & CR2 & 0.283 & 0.133 & 0.131 & 0.130\\
     100 & 0.25 & CR3 & 0.237 & 0.087 & 0.091 & 0.091\\
     100 & 0.5 & Standard & 0.877 & 0.458 & 0.452 & 0.452\\
     100 & 0.5 & CR0 & 0.802 & 0.451 & 0.454 & 0.454\\
     100 & 0.5 & CR2 & 0.753 & 0.354& 0.355 & 0.355\\
     100 & 0.5 & CR3 & 0.706 &0.271 & 0.267 & 0.267\\
     100 & 0.75 & Standard & 0.996 & 0.811 & 0.811 & 0.811\\
     100 & 0.75 & CR0 & 0.985 & 0.763 & 0.762 & 0.762 \\
     100 & 0.75 & CR2 & 0.971 & 0.662 & 0.657 & 0.657\\
     100 & 0.75 & CR3 & 0.960 & 0.550 & 0.559 & 0.559\\
     100 & 1 & Standard & 1.000 & 0.979 & 0.979 & 0.979\\
     100 & 1 & CR0 & 1.000 & 0.953 & 0.957 & 0.957\\
     100 & 1 & CR2 & 1.000 & 0.909 & 0.912 & 0.912\\
     100 & 1 & CR3 & 1.000 & 0.839 & 0.841 & 0.841\\
    \bottomrule
\end{tabular}
\end{table}
\FloatBarrier

\clearpage

\subsection{Nonlinear covariate effect}
\begin{figure}[!htb]
    \centerline{\includegraphics[width=\textwidth]{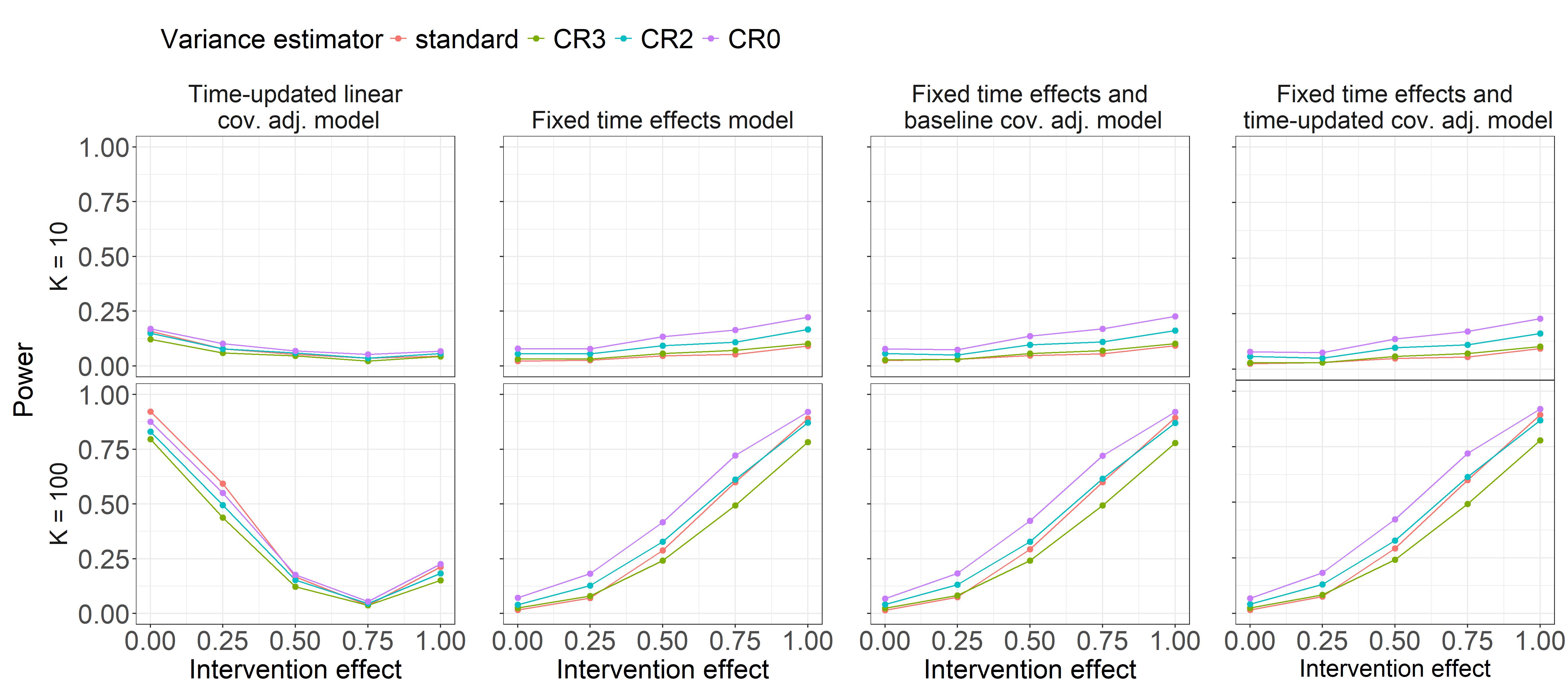}}
    \caption{Statistical power derived from four analysis model (Table~\ref{tab:mixed_models}) without (standard, red line) and with cluster-robust variance estimation, specifically CR0VE (purple line), CR2VE (blue line) and CR3VE (green line). Closed cohort data with $I = 8$ clusters, $J = 4$ steps, a cluster-period size of $K = 10$ (top) or $K = 100$ (bottom) and a nonlinear covariate effect.\label{fig:Power_closedCohort_nonlinear}}
\end{figure}
\FloatBarrier

\begin{table}[!htb]
\caption{Statistical power derived from four analysis models (Table~\ref{tab:mixed_models}) without (standard) and with cluster-robust variance estimation, specifically CR0VE, CR2VE and CR3VE. Closed cohort data with $I = 8$ clusters, $J = 4$ steps, a cluster-period size of $K = 10$ or $K = 100$ and a nonlinear covariate effect.}
\label{tab:Power_nonlinear}
\centering
\begin{tabular}{R{0.2cm}p{0.3cm}p{1.8cm}R{1.2cm}R{1.2cm}R{1.2cm}R{1.2cm}}
    \toprule
    $\mathbf{K}$ & $\boldsymbol{\theta}$ & \textbf{Variance estimator} &
    \multicolumn{4}{c}{\textbf{Model}} \\
    \cmidrule(lr){4-7}
    & & &
    \textbf{Eq. 1} &
    \textbf{Eq. 2} &
    \textbf{Eq. 3} &
    \textbf{Eq. 4} \\
    \midrule
     10 & 0 & Standard & 0.159 & 0.023 & 0.025 & 0.025\\
     10 & 0 & CR0 & 0.170 & 0.080 & 0.078 & 0.078\\
     10 & 0 & CR2 & 0.149 & 0.056 & 0.057 & 0.057\\
     10 & 0 & CR3 & 0.122 & 0.031 & 0.028 & 0.028\\
     10 & 0.25 & Standard &  0.078 & 0.026 & 0.030 & 0.030\\
     10 & 0.25 & CR0 & 0.102 & 0.078 & 0.074 & 0.074\\
     10 & 0.25 & CR2 & 0.078 & 0.055 & 0.050 & 0.050\\
     10 & 0.25 & CR3 & 0.059 & 0.032 & 0.030 & 0.030\\
     10 & 0.5 & Standard & 0.053 & 0.046 & 0.048 & 0.048\\
     10 & 0.5 & CR0 & 0.069 & 0.134 & 0.136 & 0.136\\
     10 & 0.5 & CR2 & 0.059 & 0.092 & 0.096 & 0.096\\
     10 & 0.5 & CR3 & 0.046 & 0.057 & 0.057 & 0.057\\
     10 & 0.75 & Standard & 0.035 & 0.053 & 0.055 & 0.055\\
     10 & 0.75 & CR0 & 0.053 & 0.164 & 0.170 & 0.170\\
     10 & 0.75 & CR2 & 0.036 & 0.108 & 0.110 & 0.110\\
     10 & 0.75 & CR3 & 0.023 & 0.072 & 0.070 & 0.070\\
     10 & 1 & Standard & 0.045 & 0.091 & 0.093 & 0.093\\
     10 & 1 & CR0 & 0.068 & 0.223 & 0.227 & 0.227\\
     10 & 1 & CR2 & 0.057 & 0.167 & 0.161 & 0.161\\
     10 & 1 & CR3 & 0.043 & 0.102 & 0.102 & 0.102\\
     100 & 0 & Standard & 0.922 & 0.016 & 0.015 & 0.015\\
     100 & 0 & CR0 & 0.876 & 0.072 & 0.068 & 0.068\\
     100 & 0 & CR2 & 0.830 & 0.040 & 0.041 & 0.041\\
     100 & 0 & CR3 & 0.796 & 0.025 & 0.024 & 0.024\\
     100 & 0.25 & Standard & 0.593 & 0.071 & 0.075 & 0.075\\
     100 & 0.25 & CR0 & 0.550 & 0.182 & 0.183 & 0.183\\
     100 & 0.25 & CR2 & 0.495 & 0.128 & 0.131 & 0.131\\
     100 & 0.25 & CR3 & 0.438 & 0.080 & 0.083 & 0.083\\
     100 & 0.5 & Standard & 0.168 & 0.288 & 0.293  & 0.293\\
     100 & 0.5 & CR0 & 0.177 & 0.416 & 0.423 & 0.423\\
     100 & 0.5 & CR2 & 0.153 & 0.328 & 0.328 & 0.328\\
     100 & 0.5 & CR3 & 0.122 & 0.242 & 0.242 & 0.242\\
     100 & 0.75 & Standard & 0.038 & 0.600 & 0.600 & 0.600\\
     100 & 0.75 & CR0 & 0.055 & 0.722 & 0.720 & 0.720\\
     100 & 0.75 & CR2 & 0.045 & 0.612 & 0.615 & 0.615\\
     100 & 0.75 & CR3 & 0.037 & 0.494 & 0.493 & 0.493\\
     100 & 1 & Standard & 0.212 & 0.890 & 0.894 & 0.894\\
     100 & 1 & CR0 & 0.226 & 0.920 & 0.921 & 0.921\\
     100 & 1 & CR2 & 0.183 & 0.871 & 0.870 & 0.870\\
     100 & 1 & CR3 & 0.151 & 0.783 & 0.779 & 0.779\\
    \bottomrule
\end{tabular}
\end{table}
\FloatBarrier

\clearpage

\section{Prevalence of singular fits and non-convergence}
\begin{table}[!htb]
\caption{Prevalence of singular fits and non-convergence in the time-updated linear covariate adjustment model (Equation 1) over 1,000 simulation iterations.}
\label{tab:SingFit_NonConv_Eq1}
\centering
\begin{tabular}{rrrR{0.95cm}R{0.95cm}R{0.95cm}R{0.95cm}R{0.95cm}R{0.95cm}R{0.95cm}R{0.95cm}}
    \toprule
    $\mathbf{I}$ & $\mathbf{J}$ & $\mathbf{K}$ &
    \multicolumn{2}{c}{\textbf{Closed cohort},} &
    \multicolumn{2}{c}{\textbf{Closed cohort},} &
    \multicolumn{2}{c}{\textbf{Open cohort},} &
    \multicolumn{2}{c}{\textbf{Open cohort},} \\
    & & & \multicolumn{2}{c}{\textbf{linear effect}} &
    \multicolumn{2}{c}{\textbf{nonlinear effect}} &
    \multicolumn{2}{c}{\textbf{linear effect}} &
    \multicolumn{2}{c}{\textbf{nonlinear effect}} \\
    \cmidrule(lr){4-5} \cmidrule(lr){6-7} \cmidrule(lr){8-9} \cmidrule(l){10-11}
    & & &
    \textbf{Sing. fit} &
    \textbf{Non-conv.} &
    \textbf{Sing. fit} &
    \textbf{Non-conv.} &
    \textbf{Sing. fit} &
    \textbf{Non-conv.} &
    \textbf{Sing. fit} &
    \textbf{Non-conv.} \\
    \midrule
     8 & 4 &  10 & 0.3\% & 0.4\% & 29.5\% & 40.1\% & 0.0\% & 0.1\% & 25.3\% & 31.7\% \\
    16 & 4 &  10 & 0.0\% & 0.1\% & 30.7\% & 40.8\% & 0.0\% & 0.0\% & 17.2\% & 23.6\% \\
    32 & 4 &  10 & 0.0\% & 0.1\% & 20.2\% &30.1\% & 0.0\% & 0\% & 8.7\% & 12.9\% \\
     8 & 4 & 100 & 0.0\% & 23.6\% & 6.0\% & 14.7\% & 0.0\% & 18.5\% & 1.4\% & 6.1\% \\
    16 & 4 & 100 & 0.0\% & 21.1\% & 1.4\% & 11.0\% & 0.0\% & 10.1\% & 0.1\% & 3.5\% \\
    32 & 4 & 100 &  0.0\% & 12.2\% & 0.0\% & 10.4\% & 0.0\% &  4.6\% & 0.0\% & 3.5\% \\
     8 & 8 &  10 & 0.2\% & 0.8\% & 36.1\% & 46.4\% & 0.0\% & 0.5\% & 19.2\% & 26.3\% \\
    16 & 8 &  10 & 0.0\% & 0.1\% & 29.9\% & 40.9\% & 0.0\% & 0.1\% & 10.5\% & 14.6\% \\
    32 & 8 &  10 & 0.0\% & 0.1\% & 24.6\% & 33.1\% & 0.0\% & 0.1\% & 4.0\% & 5.7\% \\
     8 & 8 & 100 & 0.0\% & 45.0\% & 6.8\% & 18.8\% & 0.0\% & 37.5\% & 0.4\% & 8.3\% \\
    16 & 8 & 100 &  0.0\% & 59.0\% & 1.3\% & 9.1\% & 0.0\% & 38.7\% & 0.0\% & 0.9\% \\
    32 & 8 & 100 &  0.0\% & 63.4\% & 0.0\% & 11.5\% & 0.0\% &  47.4\% & 0.0\% & 15.5\% \\
    \bottomrule
\end{tabular}
\end{table}
\FloatBarrier

\begin{table}[!htb]
\caption{Prevalence of singular fits and non-convergence in the fixed time effects model (Equation 2) over 1,000 simulation iterations.}
\label{tab:SingFit_NonConv_Eq2}
\centering
\begin{tabular}{rrrR{0.95cm}R{0.95cm}R{0.95cm}R{0.95cm}R{0.95cm}R{0.95cm}R{0.95cm}R{0.95cm}}
    \toprule
    $\mathbf{I}$ & $\mathbf{J}$ & $\mathbf{K}$ &
    \multicolumn{2}{c}{\textbf{Closed cohort},} &
    \multicolumn{2}{c}{\textbf{Closed cohort},} &
    \multicolumn{2}{c}{\textbf{Open cohort},} &
    \multicolumn{2}{c}{\textbf{Open cohort},} \\
    & & & \multicolumn{2}{c}{\textbf{linear effect}} &
    \multicolumn{2}{c}{\textbf{nonlinear effect}} &
    \multicolumn{2}{c}{\textbf{linear effect}} &
    \multicolumn{2}{c}{\textbf{nonlinear effect}} \\
    \cmidrule(lr){4-5} \cmidrule(lr){6-7} \cmidrule(lr){8-9} \cmidrule(l){10-11}
    & & &
    \textbf{Sing. fit} &
    \textbf{Non-conv.} &
    \textbf{Sing. fit} &
    \textbf{Non-conv.} &
    \textbf{Sing. fit} &
    \textbf{Non-conv.} &
    \textbf{Sing. fit} &
    \textbf{Non-conv.} \\
    \midrule
     8 & 4 &  10 & 32.5\% & 49.6\% & 36.2\% & 51.7\% & 32.6\% & 49.1\% & 32.5\% & 47.7\% \\
    16 & 4 &  10 & 32.8\% & 45.5\% & 31.6\% & 46.7\% & 29.3\% & 42.6\% & 28.5\% & 41.3\% \\
    32 & 4 &  10 & 25.9\% & 41.6\% & 30.7\% & 45.5\% & 27.9\% & 41.5\% & 23.1\% & 38.8\% \\
     8 & 4 & 100 & 15.7\% & 28.7\% & 17.1\% & 35.3\% & 12.8\% & 22.9\% & 11.7\% & 21.0\% \\
    16 & 4 & 100 & 10.9\% & 21.1\% & 16.1\% & 31.0\% & 6.2\% & 10.5\% & 8.2\% & 14.2\% \\
    32 & 4 & 100 &  5.0\% & 10.2\% & 10.0\% & 24.8\% & 2.0\% &  3.4\% &  3.0\% &  5.9\% \\
     8 & 8 &  10 & 29.2\% & 46.2\% & 32.2\% & 47.2\% & 32.4\% & 47.1\% & 28.6\% & 40.7\% \\
    16 & 8 &  10 & 27.8\% & 45.2\% & 31.4\% & 49.3\% & 27.4\% & 42.6\% & 28.3\% & 40.5\% \\
    32 & 8 &  10 & 26.3\% & 42.6\% & 30.5\% & 46.5\% & 26.1\% & 39.0\% & 21.0\% & 33.4\% \\
     8 & 8 & 100 & 14.4\% & 31.7\% & 15.6\% & 40.1\% & 9.7\% & 17.6\% & 9.0\% & 19.6\% \\
    16 & 8 & 100 &  7.2\% & 21.9\% & 14.3\% & 37.2\% & 3.1\% &  10.0\% &  4.6\% & 13.7\% \\
    32 & 8 & 100 &  3.5\% & 21.1\% &  0.2\% & 11.5\% & 0.3\% &  8.6\% &  0.6\% &  7.1\% \\
    \bottomrule
\end{tabular}
\end{table}
\FloatBarrier

\begin{table}[!htb]
\caption{Prevalence of singular fits and non-convergence in the fixed time effects and baseline covariate adjustment model (Equation 3) over 1,000 simulation iterations.}
\label{tab:SingFit_NonConv_Eq3}
\centering
\begin{tabular}{rrrR{0.95cm}R{0.95cm}R{0.95cm}R{0.95cm}R{0.95cm}R{0.95cm}R{0.95cm}R{0.95cm}}
    \toprule
    $\mathbf{I}$ & $\mathbf{J}$ & $\mathbf{K}$ &
    \multicolumn{2}{c}{\textbf{Closed cohort},} &
    \multicolumn{2}{c}{\textbf{Closed cohort},} &
    \multicolumn{2}{c}{\textbf{Open cohort},} &
    \multicolumn{2}{c}{\textbf{Open cohort},} \\
    & & & \multicolumn{2}{c}{\textbf{linear effect}} &
    \multicolumn{2}{c}{\textbf{nonlinear effect}} &
    \multicolumn{2}{c}{\textbf{linear effect}} &
    \multicolumn{2}{c}{\textbf{nonlinear effect}} \\
    \cmidrule(lr){4-5} \cmidrule(lr){6-7} \cmidrule(lr){8-9} \cmidrule(l){10-11}
    & & &
    \textbf{Sing. fit} &
    \textbf{Non-conv.} &
    \textbf{Sing. fit} &
    \textbf{Non-conv.} &
    \textbf{Sing. fit} &
    \textbf{Non-conv.} &
    \textbf{Sing. fit} &
    \textbf{Non-conv.} \\
    \midrule
     8 & 4 &  10 & 0.3\% & 0.4\% & 30.5\% & 41.3\% & 0.4\% & 0.6\% & 24.7\% & 33.5\% \\
    16 & 4 &  10 & 0.0\% & 0.0\% & 30.2\% & 41.9\% & 0.0\% & 0.1\% & 17.2\% & 24.5\% \\
    32 & 4 &  10 & 0.0\% & 0.1\% & 21.5\% & 30.0\% & 0.0\% & 0.0\% & 10.6\% & 13.6\% \\
     8 & 4 & 100 & 0.0\% & 24.0\% & 5.2\% & 14.9\% & 0.0\% & 14.7\% & 1.3\% & 6.0\% \\
    16 & 4 & 100 & 0.0\% & 20.5\% & 0.8\% & 11.1\% & 0.0\% & 8.7\% & 0.0\% & 3.3\% \\
    32 & 4 & 100 & 0.0\% & 11.4\% & 0.0\% & 11.2\% & 0.0\% & 4.5\% & 0.0\% & 3.0\% \\
     8 & 8 &  10 & 0.4\% & 0.9\% & 35.2\% & 46.6\% & 0.8\% & 1.5\% & 20.9\% & 27.8\% \\
    16 & 8 &  10 & 0.0\% & 0.2\% & 29.7\% & 40.1\% & 0.0\% & 0.2\% & 11.5\% & 15.2\% \\
    32 & 8 &  10 & 0.0\% & 0.0\% & 25.4\% & 35.0\% & 0.0\% & 0.0\% & 4.8\% & 7.4\% \\
     8 & 8 & 100 & 0.0\% & 45.4\% & 6.6\% & 20.8\% & 0.0\% & 23.8\% & 0.4\% & 7.8\% \\
    16 & 8 & 100 & 0.0\% & 56.1\% & 0.8\% & 9.4\% & 0.0\% & 25.6\% & 0.0\% & 10.8\% \\
    32 & 8 & 100 & 0.0\% & 66.9\% & 0.2\% & 11.5\% & 0.0\% & 33.20\% & 0.0\% & 13.8\% \\
    \bottomrule
\end{tabular}
\end{table}
\FloatBarrier

\begin{table}[!htb]
\caption{Prevalence of singular fits and non-convergence in the fixed time effects and time-updated covariate adjustment model (Equation 4) over 1,000 simulation iterations.}
\label{tab:SingFit_NonConv_Eq4}
\centering
\begin{tabular}{rrrR{0.95cm}R{0.95cm}R{0.95cm}R{0.95cm}R{0.95cm}R{0.95cm}R{0.95cm}R{0.95cm}}
    \toprule
    $\mathbf{I}$ & $\mathbf{J}$ & $\mathbf{K}$ &
    \multicolumn{2}{c}{\textbf{Closed cohort},} &
    \multicolumn{2}{c}{\textbf{Closed cohort},} &
    \multicolumn{2}{c}{\textbf{Open cohort},} &
    \multicolumn{2}{c}{\textbf{Open cohort},} \\
    & & & \multicolumn{2}{c}{\textbf{linear effect}} &
    \multicolumn{2}{c}{\textbf{nonlinear effect}} &
    \multicolumn{2}{c}{\textbf{linear effect}} &
    \multicolumn{2}{c}{\textbf{nonlinear effect}} \\
    \cmidrule(lr){4-5} \cmidrule(lr){6-7} \cmidrule(lr){8-9} \cmidrule(l){10-11}
    & & &
    \textbf{Sing. fit} &
    \textbf{Non-conv.} &
    \textbf{Sing. fit} &
    \textbf{Non-conv.} &
    \textbf{Sing. fit} &
    \textbf{Non-conv.} &
    \textbf{Sing. fit} &
    \textbf{Non-conv.} \\
    \midrule
     8 & 4 &  10 & 0.3\% & 0.4\% & 30.1\% & 41.1\% & 0.0\% & 0.1\% & 24.4\% & 33.2\% \\
    16 & 4 &  10 & 0.0\% & 0.0\% & 29.7\% & 41.4\% & 0.0\% & 0.1\% & 17.7\% & 23.6\% \\
    32 & 4 &  10 & 0.0\% & 0.1\% & 21.2\% & 30.1\% & 0.0\% & 0.1\% & 9.7\% & 12.8\% \\
     8 & 4 & 100 & 0.0\% & 23.9\% & 5.2\% & 13.9\% & 0.0\% & 16.6\% & 1.3\% & 6.9\% \\
    16 & 4 & 100 & 0.0\% & 20.4\% & 0.6\% & 11.0\% & 0.0\% & 10.1\% & 0.0\% & 2.9\% \\
    32 & 4 & 100 & 0.0\% & 11.7\% & 0.0\% & 11.4\% & 0.0\% & 4.0\% & 0.0\% & 2.5\% \\
     8 & 8 &  10 & 0.4\% & 1.0\% & 35.8\% & 46.8\% & 0.0\% & 0.5\% & 19.8\% & 26.0\% \\
    16 & 8 &  10 & 0.0\% & 0.3\% & 29.5\% & 40.2\% & 0.0\% & 0\% & 11.2\% & 15.0\% \\
    32 & 8 &  10 & 0.0\% & 0.0\% & 25.3\% & 35.1\% & 0.0\% & 0.2\% & 4.0\% & 6.6\% \\
     8 & 8 & 100 & 0.0\% & 45.1\% & 6.9\% & 19.9\% & 0.0\% & 36.1\% & 0.5\% & 7.0\% \\
    16 & 8 & 100 & 0.0\% & 55.0\% & 1.1\% & 10.5\% & 0.0\% & 38.8\% & 0.0\% & 11.1\% \\
    32 & 8 & 100 & 0.0\% & 66.1\% & 0.2\% & 11.5\% & 0.0\% &  50.4\% & 0.0\% & 15.3\% \\
    \bottomrule
\end{tabular}
\end{table}
\FloatBarrier

\end{document}